\documentclass[journal,10pt]{IEEEtran}

\usepackage[cmex10]{amsmath}
\usepackage[dvips]{graphicx}
\usepackage{ifpdf,cite,dsfont}
\usepackage{array,color,float,stfloats}
\usepackage{amssymb,subfigure,lipsum}
\usepackage{cuted,textcomp,multirow}
\usepackage{enumitem,algorithm,algpseudocode}
\usepackage{footmisc,makecell,mathtools}

\newtheorem{theorem}{Theorem}

\newtheorem{proposition}{Proposition}
\newtheorem{corollary}{Corollary}
\newtheorem{remark}{Remark}

\newcommand{\E}{\mathbb{E}}
\newcommand{\Pt}{P_\text{tx}}
\newcommand{\Nt}{\widetilde{N}}

\allowdisplaybreaks
\IEEEoverridecommandlockouts
\begin{document}

\title{Superimposed Chirp Waveforms for SWIPT with Diplexer-based Integrated Receivers}

\author{Arijit Roy, Constantinos Psomas, \IEEEmembership{Senior Member, IEEE}, and Ioannis Krikidis, \IEEEmembership{Fellow, IEEE}\vspace{-5mm}
\thanks{This work has received funding from the European Research Council (ERC) under the European Union's Horizon 2020 research and innovation programme (Grant agreement No. 819819). It was also funded by the European Union's Horizon Europe programme (ERC, WAVE, Grant agreement No. 101112697). Views and opinions expressed are however those of the author(s) only and do not necessarily reflect those of the European Union or the European Research Council Executive Agency. Neither the European Union nor the granting authority can be held responsible for them.}% \textit{(Corresponding author: Arijit Roy.)}
\thanks{The authors are with the Department of Electrical and Computer Engineering, University of Cyprus, Nicosia 1678, Cyprus (e-mail: roy.arijit@ucy.ac.cy; psomas@ucy.ac.cy; krikidis@ucy.ac.cy).}}

\maketitle

\begin{abstract}  
In this paper, we present the superposition of chirp waveforms for simultaneous wireless information and power transfer (SWIPT) applications. Exploiting the chirp waveform characteristics enables us to superimpose multiple chirps, thereby allowing transmission of the same number of waveforms over less bandwidth. This enables us to perform subband selection when operating over set of orthogonal subbands. Furthermore, we consider a user equipped with a diplexer-based integrated receiver (DIR), which enables to extract radio frequency power and decode information from the same signal without splitting. Thereby, incorporating chirp superposition and subband selection, a transmission scheme is proposed to exploit both the diode's nonlinearity and frequency diversity. We derive novel closed-form analytical expressions of the average harvested energy (HE) via transmission of superimposed chirp over selected subbands based on tools from order statistics. We also analyze the downlink information rate achieved at the user. Through our analytical and numerical results, for the considered system setup, we show that superimposed chirp-based SWIPT provides an improvement of 30$\%$ in average HE performance as compared to multisine waveforms consisting of a set of fixed-frequency cosine signals, improves the minimum level of HE in a multiuser network, and extends the operating range of energy transfer as compared to fixed-frequency waveforms. Furthermore, we illustrate that the inclusion of DIR at the receiver for SWIPT enlarges the energy-information transfer region when compared to the widely considered power splitting receiver. 
\end{abstract}


%\newpage
\section{Introduction}\label{sec:Introduction}
\IEEEPARstart{T}{ransfer} of wireless energy through radio frequency (RF) signals allows to prolong the operational lifetime of a user terminal (UT) in an energy-constrained wireless network (WN)~\cite{Popovic}. Particularly, in scenarios where users might be located at hazardous, inaccessible, and/or inconvenient locations to recharge or replace the batteries. Wireless power transfer (WPT) is a promising energy harvesting technology as it can provide a sustainable and predictable power supply compared with other energy sources \cite{Mukherjee}. Moreover, it holds enormous potential to power a massive number of Internet-of-Things (IoT) UTs for applications like massive machine-type communications in next generation (beyond 5G and 6G) wireless networks \cite{Chen, Cisco}. The WPT architecture at the users generally considers either a simplified linear energy harvesting model \cite{Zhou2}, or a more practical nonlinear energy harvesting model based on the nonlinearities of the rectifier \cite{Clerckx2}. Furthermore, the UTs in the network can communicate over the same RF signal to be using for power transmission. Thereby, simultaneous wireless information and power transfer (SWIPT) has attracted significant attention in the research community \cite{Krikidis,Wang}. To realize SWIPT, two receiver architectures, based on the processing of the received signal, are considered, i.e., the separated receiver and the integrated receiver \cite{Zhou}. The separated receiver architecture broadly considers two schemes, namely, power splitting (PS) and time switching (TS) \cite{Zhou2}. In PS, the received signal is split into two parts for energy harvesting and information decoding. In TS, the received signal is processed by an energy receiver or an information receiver at a time. On the other hand, in an integrated receiver based on the diplexer \cite{Besma}, studied recently, the entire received signal passes through the rectifier and frequency-domain multiplexing through an RF diplexer is performed to extract RF power and information from the same signal. 

In recent studies, improving the efficiency and overcoming the limitations of SWIPT (and WPT) systems has been explored. The work in \cite{Zhou} investigates the energy-information rate trade-off in a single-carrier SWIPT system based on a dynamic PS scheme assuming perfect channel state information (CSI). In \cite{Feng}, the authors studied a three-node wireless powered communication system and obtained optimal time allocation for the power transfer and information transfer phases to maximize the system throughput. The authors in \cite{Zhao}, present a SWIPT system to maximize the harvested energy (HE) by the users while maintaining a target data rate for information users. Transmission of information and energy in a massive multiple input multiple-output (MIMO) system is presented in \cite{Long} where the receivers use either PS or TS schemes with the aim of either maximizing the minimum transmission rate or optimizing the energy efficiency of the system. The design of a massive MIMO SWIPT system to obtain the max-min rate-energy region for TS and PS protocols is presented in \cite{Kudathanthirige}. For a time-slotted SWIPT system the authors in \cite{Morsi} study the transmit waveform optimization to maximize the energy-information rate region in a network with both information and energy users.

The authors in \cite{Ng} consider the PS protocol and assume perfect CSI to optimize the energy efficiency of a SWIPT network over orthogonal frequency division multiplexing (OFDM) system for a single user. For an OFDM based system with perfect CSI, the transmit power policy was obtained to maximize the sum data rate for the TS receivers subject to a HE constraint in \cite{Zhou2}. For a single user system, a two phase channel training protocol over OFDM was presented in \cite{Zeng} to maximize the HE through subband selection. Assuming perfect CSI, transmission of wireless information and power to a single user over OFDM system to maximize the HE was considered in \cite{Lu}, where the user is assumed to have a set of bandpass filters for extracting the required RF signals to perform both information decoding and energy harvesting operations. The authors in \cite{Clerckx1} optimize the amplitudes and phases of multisine waveforms \cite{Boaventura} as a function of the subband channel response to improve the WPT efficiency of a multiuser WPT system by assuming perfect CSI. In \cite{Chaehun}, the authors proposed a multi-tone amplitude modulation and associated receiver architecture for SWIPT to improve the symbol error rate (SER) performance and energy efficiency while reducing the receiver complexity. Transmission of signals over all available subbands (with adaptive amplitude and phase) in a multiband system is beneficial due to the nonlinearity of the diode and improves the output power level \cite{Clerckx1}. On the other hand, \cite{Zeng} suggests the transmission of power over selected subbands having relatively higher channel gains, thereby limiting the number of transmitted signals, to attain higher frequency diversity gain and improving the HE. This leads to the question on how to exploit the benefits of both these techniques, i.e., diode nonlinearity gain and frequency diversity gain. Furthermore, unlike most of the works on SWIPT where the UT either considers separate energy and information receivers or splits the received signal (immediately after reception or at the output of the low-pass filter (LPF)) for energy harvesting and information decoding operations, respectively, an RF diplexer-based integrated receiver (DIR)~\cite{Besma} enables to extract both RF power and data from the same received signal which allows to improve the HE. 

Motivated by these, we investigate the feasibility of SWIPT based on the transmission of chirp waveforms with a DIR at the UT. A chirp waveform is a frequency modulated signal, which has the following properties: (i) the duration of a chirp can be varied independently of its bandwidth, unlike fixed-frequency waveforms where bandwidth is inversely related to its duration, and (ii) the instantaneous frequency of a chirp waveform varies as a function of time~\cite[Ch.\ 6]{Cook}. Note that, the diode nonlinearity gain is proportional to the number of distinct waveforms transmitted~\cite{Clerckx1}. Therefore, given a set of subbands $N$, with conventional fixed-frequency cosine waveforms, one can transmit a set of maximum $N$ waveforms. This means the transmission over all available subbands, irrespective of their channel gains. On the other hand, frequency diversity gain is obtained through the transmission over a set of subbands with relatively higher channel gains~\cite{Zeng}. As a chirp waveform can handle the time and frequency in a more flexible way, by exploiting the properties of chirp, we present the design of superimposed chirps and an associated transmission strategy for SWIPT to attain jointly frequency diversity and diode nonlinearity gains. Specifically, we propose the superposition of multiple distinct chirp waveforms and their transmission over selected subbands with relatively higher channel gains which leads to frequency diversity gain. Moreover, through the superposition of multiple chirps in a given subband, we are able to use a smaller set of subbands with higher channel gains while maintaining the total number of waveforms equal to the available subbands $N$. In this way, we can achieve more degrees of freedom in terms of selection of bands with higher gains without impacting the nonlinearity gains. Therefore, chirp waveform provides unique benefits over conventional fixed-frequency waveforms. Furthermore, we also focus on ascertaining whether employing superimposed chirp with DIR at the user could substantially improve the level of HE while maintaining a reasonable information rate at the UT. Note that the construction of superimposed chirp is different from the design of basic chirp that finds wide applications in radar, communication, and design of LoRa-based modulation for IoT wide area networks \cite{Cook,Roy,Elshabrawy}. 

In contrast to existing works, to the best of our knowledge, this is the first work that \textbf{(i)} considers the design and application of superimposed chirp waveforms for SWIPT, \textbf{(ii)} combines superimposed chirp along with a diplexer based integrated receiver, \textbf{(iii)} analyzes and quantifies the gains that can be obtained in HE when superimposed chirp is transmitted over subbands selected independently for each user and with estimated CSI in a multiuser scenario. Specifically, we make the following key contributions: 
\begin{itemize}
\item 
We present the design of superimposed chirp waveforms and we investigate a transmission strategy based on the selection of subbands via estimated CSI for downlink (DL) SWIPT to achieve the benefits of both diode nonlinearity gain and frequency diversity gain, in a multiuser network. We demonstrate that the proposed scheme achieves a notable improvement in the HE and the operational range of energy transfer over fixed-frequency waveforms.	
	
\item 
Based on the transmission scheme, we derive novel analytical expressions of the average HE for superimposed chirp at the DIR enabled users. The mathematical analysis takes into account the nonlinearity of the harvester and exploits the order statistics of the subband channel estimates. We also derive the associated expression for the average information received by the user. As a benchmark and a special case, we present the expressions for the case of fixed-frequency waveform also. The results indicate the joint impact of different design parameters on the HE and information rate of the SWIPT system. 
		
\item 	
We investigate the performance of a conventional PS receiver for SWIPT with superimposed chirp and proposed transmission scheme. The outcomes are compared against its DIR counterpart. We demonstrate that DIR outperforms the PS receiver in terms of HE while maintaining a certain information rate.
\end{itemize}
The rest of this paper is organized as follows: Section II presents the system model, preliminaries on chirp and DIR, and the channel estimation. Section III includes the design of superimposed chirp, the transmission strategy and the DL beamforming for SWIPT, as well as the analysis of average HE and information rate for DIR. Numerical results are presented in Section IV, followed by the conclusions in Section V.

\textit{Notation:} Boldface lower-case letters and upper-case letters denote vectors and matrices, respectively. Scalars are denoted by letters (or Greek letters). $\mathbf{D}$ stands for a diagonal matrix. $\mathbf{I}_N$ indicates an $N\times N$ identity matrix. $\E\{\cdot\}$ is the expectation operation. $(\cdot)^\ast$, $(\cdot)^\top$, and $(\cdot)^\text{H}$ denotes the conjugate, the transpose, and the conjugate transpose of a vector or matrix, respectively. $\|{\cdot}\|$ denotes the Euclidean norm. Distribution of a circularly symmetric complex Gaussian (CSCG) random vector $\mathbf{x}$ with mean $\boldsymbol{m}$ and covariance matrix $\mathbf{C}$ is denoted by $\mathbf{x} \sim \mathcal{CN}(\boldsymbol{m},\mathbf{C})$. Given a complex number $z$, $\Re(z)$ and $\Im(z)$ indicate the real and imaginary part, respectively. $F_A^{(l)}(a,\mathbf{b};\mathbf{c};\mathbf{z})$ is the Lauricella function of type A. $\gamma(\cdot,\cdot)$ is the lower incomplete gamma function.

\section{System Model}
\begin{figure}[!t]\centering
\subfigure[]{\includegraphics[width=0.52\linewidth]{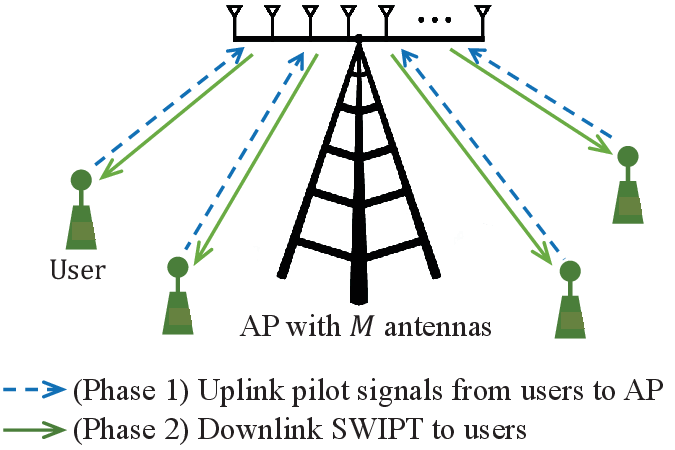}}\hfill
\subfigure[]{\includegraphics[width=0.48\linewidth]{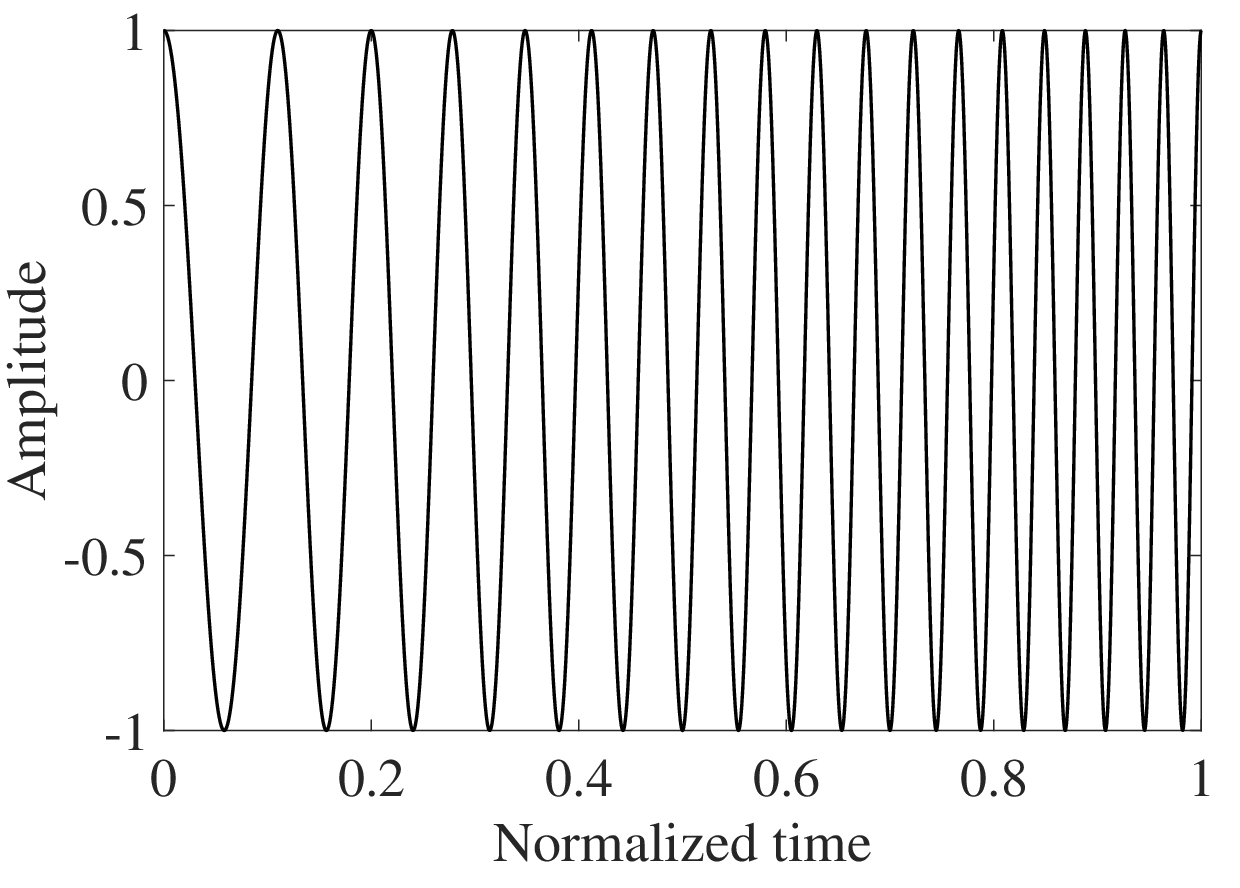}}
\caption{(a). The considered system model. (b) An up-chirp waveform.}
\label{fig:system_model_sweit}\vspace{-4mm}
\end{figure}

We consider an OFDM-based system comprising of an access point (AP) with $M$ antennas performing DL SWIPT to $K$ single-antenna UTs, as shown in Fig. \ref{fig:system_model_sweit}(a). The total system bandwidth is equally divided into $N$ subbands with subband bandwidth $B \!\leq\! B_{\text{ch}}$, where $B_{\text{ch}}$ is the coherence bandwidth. The channel over each subband experiences frequency-flat fading and varies independently across subbands \cite{Zhou2, Zeng}. The complex baseband channel vector between the AP and the $k^\text{th}$ user over the $n^\text{th}$ subband is
\begin{align}
\mathbf{g}_{n,k} = \sqrt{\beta_k} \mathbf{h}_{n,k} \in \mathbb{C}^{M\times 1},	
\end{align}
where $\beta_k$ denotes large-scale fading comprising of distance-dependent path-loss, $\mathbf{h}_{n,k}$ denotes small-scale fading and is a random vector with independent and identically distributed (i.i.d.) zero mean unit variance CSCG entries, i.e., $\left[\mathbf{h}_{n,k} \right]^m = h_{n,k}^m \!\sim\! \mathcal{CN}(0,1)$ \cite{Yang}, where $m=1,2,\ldots,M$, $n=1,2,\ldots,N$, and $k=1,2,\ldots,K$. 

We assume that the channel remains time-invariant over a coherence block of length $\tau_{\text{ch}}$ seconds and varies independently across blocks. Therefore, channel estimation must be done at each coherence block. Moreover, we consider that the system operates in a time-division-duplexing (TDD) mode so that the DL channel from the AP to the users is the same as the uplink (UL) channel from the users to the AP. In each interval of length $\tau_{\text{ch}}$, the communication between the AP and the users takes place in two different phases as shown in Fig.~\ref{fig:system_model_sweit}(a). In the first phase, every user transmits a pilot of length $\tau_\text{p}$ to the AP for estimating the channel impulse response. In the second phase of length $\tau_\text{DL}$ the AP beamforms on the DL for SWIPT to the users based on the estimated channel and exploiting channel reciprocity. Note that $\tau_{\text{ch}}= \tau_\text{p}+\tau_\text{DL}$. In addition, we consider that the AP is subject to an average transmit power constraint of $\Pt$. 

\subsection{Chirp Waveforms}
A chirp is a waveform in which the instantaneous frequency varies as a function of time, and referred to as up-chirp (frequency increases with time) and down-chirp (frequency decreases with time). An up-chirp waveform with bandwidth $B$ and duration $T$ over subband $n$ can be expressed as \cite{Roy}
\begin{equation}
s_n(t) = \cos \left[2\pi t \left(f_n + \frac{\mu}{2} t \right) \right], \quad 0 \leq t \leq T,\label{eq:signal_chirp}
\end{equation}
where $f_n = f_0 + (n-1)B$, $n\in \{1, 2, \dots, N\}$, and $f_0$ is the initial frequency. The instantaneous frequency $f_n + \mu t$ indicates that it varies with time. The associated rate of change of the instantaneous frequency is given by the chirp rate $\mu=\frac{B}{T}$ with $BT \geq 1$ \cite{Cook}. An up-chirp waveform is shown in Fig. \ref{fig:system_model_sweit}(b). An interesting property of chirp is that by selecting an appropriate value for $\mu$, one can vary the duration of a chirp independently of its waveform bandwidth (unlike a fixed-frequency waveform). Alternatively, given a specific bandwidth, one can choose the duration as required, and calculate the value of $\mu$ accordingly. This gives a powerful leverage over fixed-frequency in terms of designing waveforms for SWIPT, as presented in the next section.\vspace{-1mm}

\subsection{Diplexer-based Integrated Receiver}
\begin{figure}[!t]\centering
	\includegraphics[width=0.9\linewidth]{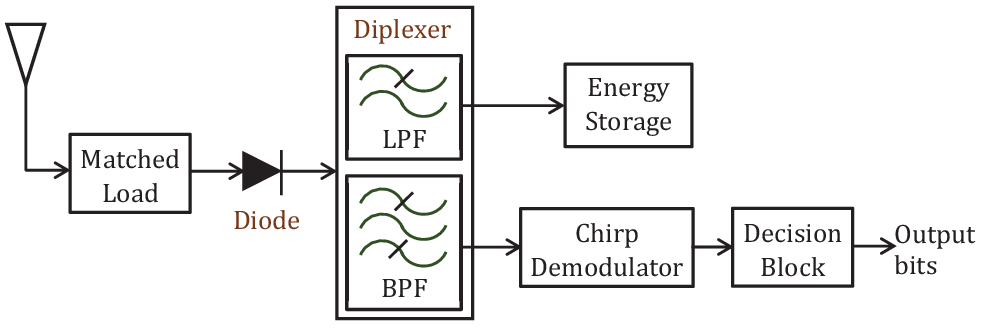}\vspace{-1mm}
	\caption{Block diagram of the diplexer-based integrated receiver.}
	\label{fig:integrated_rx_sweit}\vspace{-4mm}
\end{figure}
We consider a diplexer-based unified receiver model, based on the design in~\cite{Besma}, as shown in Fig. \ref{fig:integrated_rx_sweit}. The model in~\cite{Besma} refers to a general circuit and can be used for any waveform. Following this fact, we have focused on the application of chirp waveforms for SWIPT due to their unique properties, and the DIR circuit  provides the ability to extract both RF power and data from the same received signal without splitting the resources.
	
In general, an energy harvesting receiver is equipped with an antenna followed by a rectifier, which generally consists of a diode (e.g., a Schottky diode) together with a low LPF. In the considered DIR, an RF diplexer replaces the LPF. The diplexer allows frequency-domain multiplexing by separating the low and high frequency signals from the same input signal without splitting. Therefore, in the context of SWIPT, the low-pass band-pass diplexers can be used to separate the low frequency signals for energy harvesting operation while the higher frequency signals for information decoding~\cite{Besma}. For an RF input signal, a nonlinear device, e.g. diode, produces spectral components such as direct current (DC), fundamental frequency, harmonics of the fundamental, and intermodulation mixing products at its output~\cite{Besma}. The diode output current is processed by the duplexer low-pass filter that removes the high-frequency components, and the DC signal component available at the output of the LPF is stored in a battery or used to perform tasks related to sensing, computing and communicating. On the other hand, the output of the band-pass filter (BPF) is fed to a chirp correlator that performs multiplication of the input signal with respective reference chirp followed by integration (performed inside chirp demodulator block)\footnote{Demodulation of superimposed chirps is described in Section III.D.\vspace{-0mm}}. Then, the correlator output is fed to a decision block for the estimation of the transmitted bits through the chirps over selected subbands. %\vspace{-1mm}

\subsection{UL Pilot Transmission and Channel Estimation}
The pilot signal $s_{\pmb{\phi}_{n,k}}(t)$ transmitted over the $n^\text{th}$ subband by the $k^\text{th}$ user is given by
\begin{equation}
s_{\pmb{\phi}_{n,k}}(t) = \sum_{p=1}^{\epsilon} \pmb{\phi}_k[p] s_n(t-(p-1)T_\text{o}),
\label{eq:pilot-signal1}
\end{equation}
where $\epsilon \geq K$, $\pmb{\phi}_k \in \mathbb{C}^{1\times \epsilon}$ denotes $\epsilon$-length pilot sequence assigned to the $k^{\text{th}}$ user and $s_n(t)$ is the equivalent complex form of \eqref{eq:signal_chirp}, given by $s_n(t) = e^{j 2 \pi t\left[f_0 +(n-1)B +\frac{1}{2}{\mu}t \right]}$, $0 \leq t \leq T_\text{o}$, having unit average power and of duration $T_\text{o}=\frac{1}{B}$ over the $n^\text{th}$ subband. The duration $\tau_\text{p}$ of the pilot signal $s_{\pmb{\phi}_{n,k}}(t)$ equals $\tau_\text{p}=\epsilon T_\text{o} < \tau_\text{ch}$. The pilot sequences assigned to the $K$ users are chosen such that $\frac{1}{\epsilon} \pmb{\phi}_k \pmb{\phi}_{k^\prime}^{\text{H}}= 1$ for $k = k^\prime$ and is equal to zero otherwise, to ensure that assigned pilot sequences are mutually orthogonal.

Then, the signal received at the AP over the $n^{\text{th}}$ subband is given by
\begin{align}
\mathbf{y}_{\text{p}_n}(t) = \sqrt{\frac{P_\text{p}}{N}} \sum_{k^\prime = 1}^K \mathbf{g}_{n,k^{\prime}} s_{\pmb{\phi}_{n,k^{\prime}}}(t) + \mathbf{w}_{\text{p}_n}(t),
\end{align}
where $0 \leq t \leq \epsilon T_\text{o}$, $P_\text{p}$ denotes the pilot power and $\mathbf{w}_{\text{p}_n}(t)$ denotes the thermal noise vector at the AP. For the estimation of $\mathbf{g}_{n,k}$, we perform 
\begin{align}
\mathbf{y}_{\text{p}_{n, k}} &= \frac{1}{\sqrt{\epsilon T_\text{o}}}\int_0^{\epsilon T_\text{o}} \mathbf{y}_{\text{p}_{n}} (t) s_{\pmb{\phi}_{n,k}}^{\text{H}}(t) dt \nonumber\\
&= \sqrt{\frac{E_\text{p}}{N}} \mathbf{g}_{n,k} + \mathbf{w}_{\text{p}_{n, k}},
\label{eq:sufficient statistics}
\end{align}
where $\mathbf{y}_{\text{p}_{n, k}} \!\in\! \mathbb{C}^{M \times 1}$, $\int_{0}^{\epsilon T_\text{o}}s_{\pmb{\phi}_{n,k^{\prime}}}(t) s_{\pmb{\phi}_{n,k}}^{\text{H}}(t) dt = \epsilon T_\text{o}$ if $k \!=\! k^{\prime}$ and equal to zero otherwise, $E_\text{p} \!=\! P_\text{p} \epsilon T_\text{o}$ is the energy allocated to the UL pilot transmission, and $\mathbf{w}_{\text{p}_{n, k}} \!\in\! \mathbb{C}^{M \times 1}$ is CSCG noise,
$\mathbf{w}_{\text{p}_{n, k}} \sim \mathcal{CN}(\pmb{0}, \sigma^2 \mathbf{I}_M)$, where $\sigma^2$ denotes the noise power spectral density. Based on $\mathbf{y}_{\text{p}_{n, k}}$ in \eqref{eq:sufficient statistics}, the minimum mean square error (MMSE) estimate $\widehat{\mathbf{g}}_{n,k}$ of the channel vector $\mathbf{g}_{n,k}$ \cite{Marzetta}%\vspace{-1mm}
\begin{equation}
\widehat{\mathbf{g}}_{n,k} = \mathbf{g}_{n,k} + \widetilde{\mathbf{g}}_{n,k},
\label{eq:estimated_channel_vector}
\vspace{-1mm}
\end{equation}
where $\widetilde{\mathbf{g}}_{n,k} \sim \mathcal{CN}(\mathbf{0},(\beta_k - \gamma_k) \mathbf{I}_M)$ and
\begin{align}
\gamma_k = \frac{\frac{E_\text{p}}{N} \beta_k^2 }{\sigma^2 +\frac{E_\text{p}}{N}\beta_k},
\end{align}
and $\widetilde{\mathbf{g}}_{n,k}$ denotes the channel estimation error, uncorrelated to $\widehat{\mathbf{g}}_{n,k}$.

\begin{figure}[!t]\centering \vspace{-2mm}
	\subfigure[]{\includegraphics[width=0.43\linewidth]{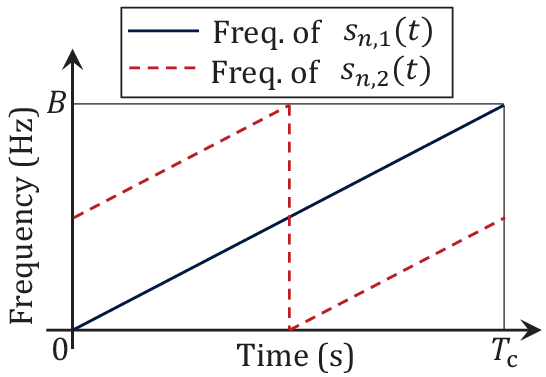}}\hfill
	\subfigure[]{\includegraphics[width=0.57\linewidth]{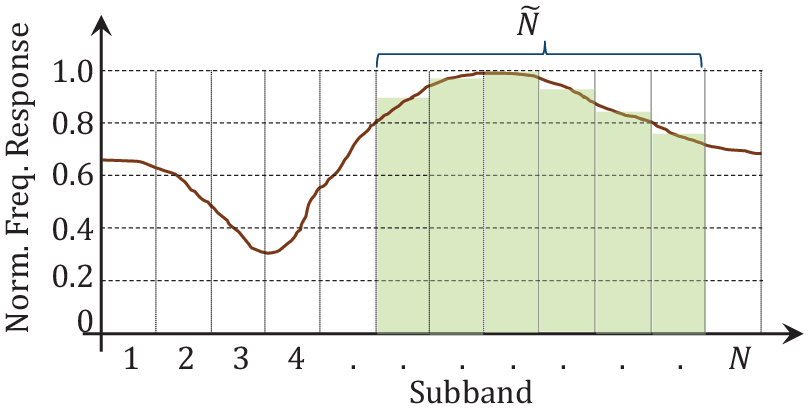}}\vspace{-1mm}
	\caption{(a). Instantaneous frequency of chirp waveforms in chirp-2 configuration. (b) Selection of $\Nt$ subbands with power allocation strategy.} \label{fig:dual_chirp_sc_selec}\vspace{-2mm}
\end{figure}

\section{Superimposed Chirp-Based SWIPT }
In this section, we present the design of superimposed chirp waveform and transmission strategy for SWIPT. We evaluate the average HE and information rate attained by a UT equipped with DIR in a multiuser network.  As a special case, the associated analysis for fixed-frequency waveform is also performed.

\subsection{Design of Superimposed Chirp Waveform}
By exploiting the chirp properties, we perform the transmission of a superposition of multiple chirp waveforms in a given subband. Defining the configuration factor $\xi$ as the number of chirps that can be superimposed within the same time and signal bandwidth of $B$, the chirp waveform over subband $n$ can be expressed as
\begin{align}
\!\!s_{n,l}(t) \!=\! \begin{cases}
\!\cos\!\left[2\pi t\! \left(\!f_n + \frac{l-1}{\xi}B +\frac{1}{2}{\mu}t \right)\! \right]\!\!, 0 \leq t \leq T_l,\!\!\\
\!\cos\!\left[2\pi t\!\left(\!f_n + \frac{l-1-\xi}{\xi}B +\frac{1}{2}{\mu}t \right)\! \right]\!\!, T_l < t \leq T,\!\!\!
\end{cases}
\label{eq:signal_xi-chirp}
\end{align}
where $\xi$ is a positive integer, $l \in \{1,2,\ldots,\xi\}$, $\mu = \frac{B}{T}$, $T_l = T \left(1-\frac{l-1}{\xi} \right)$, $T_\text{o}=\frac{1}{B}$, and $T = \xi T_\text{o}$. Considering a given peak transmit power constraint over a waveform, the waveforms in \eqref{eq:signal_xi-chirp} satisfy the constraint, since the instantaneous frequency corresponding to these superimposed waveforms appears at distinct time instants. Specifically, the separation in instantaneous frequency between any two adjacent chirps in the set of $\xi$ waveforms is $B/\xi$, while each of the waveform maintaining a bandwidth of $B$.\footnote{Note that we cannot transmit a superposition of multiple fixed-frequency waveforms, each of bandwidth $B$, within a subband of width $B$ while maintaining the peak power constraint or the orthogonality among the waveforms.} Thereby, at a given time the power contained at any frequency of a chirp remains within the given peak power level irrespective of the number of chirps that are superimposed within each subband. For example, the instantaneous frequency variation of a set of two superimposed chirp waveforms $(\xi=2)$, namely `chirp-2' is shown in Fig.~\ref{fig:dual_chirp_sc_selec}(a). Note that the waveforms in \eqref{eq:signal_xi-chirp} maintain orthogonality across subbands, and remain uncorrelated in a subband. 

\subsection{Integrated DL Energy and Information Beamforming}
We propose the following transmission strategy for DL-SWIPT based on superimposed chirp in order to beamform energy and information towards the $K$ users simultaneously. 

For the system under consideration, frequency diversity is essential, and can be attained by selecting channels with higher gains. Therefore, to get the best channels for transmission, the estimated channel vectors of each user over all subbands are arranged in descending order of their channel gains. Specifically, corresponding to the $k^{\text{th}}$ user, we sort its subband channel gains as
\begin{equation}
{\left\|\widehat{\mathbf{g}}_{[1],k}\right\|}^2 \geq {\left\|\widehat{\mathbf{g}}_{[2],k}\right\|}^2 \geq  \ldots \geq {\left\|\widehat{\mathbf{g}}_{[n],k}\right\|}^2 \geq \ldots \geq {\left\|\widehat{\mathbf{g}}_{[N],k}\right\|}^2,
\end{equation}
where $\widehat{\mathbf{g}}_{[n],k}$ denotes the estimated channel vector between the $M$ antennas of the AP and the $k^\text{th}$ user over its $n^\text{th}$ best subband and
\begin{equation}
{\left\|\widehat{\mathbf{g}}_{[n],k}\right\|}^2 = \big|{\widehat{g}_{[n],k}^1}\big|^2 + \big|{\widehat{g}_{[n],k}^2}\big|^2 + \ldots + \big|{\widehat{g}_{[n],k}^M}\big|^2.
\end{equation}
Based on this, the \textit{AP selects the $\Nt$ best subbands out of the available $N$ subbands for each of the $K$ users independently}. 
It then performs \textit{power allocation in proportion to the channel gains over the selected subbands of the $K$ users} such that the transmit power $\Pt$ constraint is satisfied with equality, as shown in Fig. \ref{fig:dual_chirp_sc_selec}(b), where the power allocated for a signal over the $n^{\text{th}}$ best subband of the $k^{\text{th}}$ user is given by $P_{[n],k}$. Note that, the subband index of the $1^{\text{st}}$ best subband, i.e., $[n = 1]$ (and all other ordered subbands as well) can be any value between $1$ to $N$. And, for each user, the mapping between the ordered subbands and their corresponding subband indices is stored in a look-up table at the AP. Combining superimposed chirp with subband selection and beamforming enables to simultaneously incorporate the following gains: frequency diversity gain, beamforming gain, and gain related to the rectifier's nonlinearity.

To elucidate further, given a channel frequency response to the $k^{\text{th}}$ user over its $N$ subbands, as depicted in Fig. \ref{fig:dual_chirp_sc_selec}(b), with superimposed chirp having configuration factor $\xi$, the AP selects the top $\Nt=\frac{N}{\xi}$ subbands having relatively higher channel gains over which it performs DL transmission; note that $\xi$ is an integer multiple of $\Nt$. The AP uses the look-up table to find the subband indices corresponding to those top $\Nt$ ordered subbands for DL transmission. As the number of transmitted chirps corresponding to each user remains equal to $N=\xi \Nt$, thus retain the nonlinearity gain (similar to that of transmitting single waveform over $N$ bands). On the other hand, transmission of superimposed chirp over $\Nt$ subbands can be interpreted as transmission of $\xi$ set of $\Nt$ waveforms. Furthermore, the power of the signals is in proportion to channel gain. Therefore, as the signal is transferred over channels having relatively higher gains, the accumulated gain for $\xi$ set will be higher as compared to the conventional scenario where signal is allocated over all $N$ subbands. Thereby, \textit{transmission of superposition of chirp waveforms over each of the} $\Nt$ \textit{subbands having relatively higher channel gains allows to attain both frequency diversity and diode nonlinearity gains}. This will result in higher received energy at the UTs which leads to a higher HE as compared to that with fixed-frequency waveform of bandwidth $B$. 

For SWIPT based on superimposed chirp waveforms with configuration factor $\xi$, let $\Xi_k$ denote the set of subband entries of the $\Nt$ best subbands for the $k^\text{th}$ user. In what follows, we consider binary phase-shift keying (BPSK) modulation for data transmission but higher order modulations, e.g., QPSK, QAM, could be also used. Therefore, the modulated signal transmitted from the AP to $K$ users over the $n^\text{th}$ subband in one symbol period $T$ is
\setcounter{equation}{10}
\begin{equation}
\label{eq:Tx_chirp_Downlink_ecsi}
\mathbf{x}_n(t) = \sqrt{\Pt} \sum_{k=1}^{K} \sqrt{\eta_{n,k}} {\pmb{\varphi}_{n,k}^\ast} S_{n,k}(t) \mathds{1}_{\{n,k\} \in \Xi_k},
\end{equation}
where $0\leq t\leq T$, $T=\xi T_\text{o}$, subband index $n \in \{1,2,\ldots,N\}$, and $\pmb{\varphi}_{n,k}$ is the beamforming precoder given by
\begin{equation}
\pmb{\varphi}_{n,k} = \frac{\widehat{\mathbf{g}}_{n,k}}{\big\|{\widehat{\mathbf{g}}_{n,k} }\big\|},\label{eq:beamform_Downlink_ecsi}
\end{equation}
and
\begin{equation}
S_{n,k}(t) = \sqrt{2} \sum_{l=1}^\xi d_{n,k,l} s_{n,l}(t),
\end{equation}
where $s_{n,l}(t)$ is given by \eqref{eq:signal_xi-chirp}, $d_{n,k,l} \in \{\pm b\}$ are the data symbols considered i.i.d. normal random variables with zero-mean and unit variance, i.e., $d_{n,k,l} \sim \mathcal{N}(0,1)$ and the scaling by $\sqrt{2}$ ensures that the average power of the chirp waveform over the $n^\text{th}$ subband is unity. Note that $\mathds{1}_{\{n,k\} \in \Xi_k}$ is an indicator function that takes the value $1$ if the $n^{\text{th}}$ subband of the $k^{\text{th}}$ user belongs to the set $\Xi_k$ of the entries of the $\Nt$ best subbands corresponding to the $k^\text{th}$ user. The power control coefficients $\eta_{n,k}$ must be chosen such that the average transmit power constraint is satisfied at the AP, i.e., $\sum_{n=1}^N \E\{\left\|\mathbf{x}_n(t) \right\|^2\} \leq \Pt$. To ensure this, we show in Appendix \ref{app:Tx_power_constraint_chirp_SWEIT} that $\sum_{k=1}^K \sum_{n=1}^N \eta_{n,k} \leq 1/\xi$.

\begin{remark}
For transmission based on fixed-frequency waveform, the modulated signal transmitted from the AP to the $K$ users over the $n^\text{th}$ subband in one symbol period $T_\text{o}$ is given by
\begin{equation}
\mathbf{x}^\prime_n(t) = \sqrt{\Pt} \sum_{k=1}^{K} \sqrt{\eta_{n,k}} {\pmb{\varphi}_{n,k}^\ast} S^{\prime}_{n,k}(t) \mathds{1}_{\{n,k\} \in \Xi_k}, 	
\label{eq:Tx_ofdm_Downlink_ecsi}
\end{equation}
where $0\leq t\leq T_\text{o}$, $S^{\prime}_{n,k}(t)= d_{n,k}s_n(t)$, $d_{n,k} \sim \mathcal{N}(0,1)$, $s_n(t)= \sqrt{2} \cos[2\pi \left(f_0 +(n-1)B \right)t]$, $n \in \{1,2,\ldots,N\}$, $T_\text{o}=\frac{1}{B}$, and $\Xi_k$ holds the ordered set of subband entries of the $N$ subbands for the $k^\text{th}$ user. Note that when transmission over all $N$ subbands is performed without selection, $\mathds{1}_{\{n,k\} \in \Xi_k}=1, \forall n,k$. Furthermore, the power control coefficients in this case must satisfy $\sum_{k=1}^{K} \sum_{n=1}^{N} \eta_{n,k} \leq 1 $ to ensure that $\sum_{n=1}^{N} \E\left\{\left\|\mathbf{x}^\prime_n(t) \right\|^2\right\} \leq \Pt$.
\label{downlink_beamform_ofdm}
\end{remark}

\subsection{Average Harvested Energy} 
\label{subsec:Avg_rx_energy_chirp}

\setcounter{equation}{15}
\begin{figure*}[t!]\vspace{-4mm}		
	\begin{align}\label{eq:total_received_signal_chirp}
	y_k(t) &= \sum_{n=1}^N \!\Bigg(\! \sqrt{\Pt \eta_{[n],k}} {\mathbf{g}_{[n],k}^\top} {\pmb{\varphi}_{[n],k}^\ast}
	S_{[n],k}(t) +\! \sum_{\substack{j=1; j\neq k}}^K \sqrt{\Pt\eta_{[n],j}} {\mathbf{g}_{[n],k}^\top} {\pmb{\varphi}_{\{[n],k\},j}^\ast} S_{\{[n],k\},j}(t) {\mathds{1}_{\{[n],k\} \in \Xi_j}} \!+ w^A_{[n],k} (t) \Bigg)\!. \vspace{-2mm}\\		
	y_k(t) &= \sum_{n=1}^{\Nt} \Bigg(\sqrt{\Pt\eta_{[n],k}} {\mathbf{g}_{[n],k}^\top} {\pmb{\varphi}_{[n],k}^\ast}
	S_{[n],k}(t) + \sum_{j=1; j\neq k}^K \sqrt{\Pt\eta_{[n],j}} {\mathbf{g}_{\{[n],j\},k}^\top} {\pmb{\varphi}_{[n],j}^\ast} S_{[n],j}(t)\Bigg) + \sum_{n=1}^N w^A_{[n],k} (t).\label{eq:total_received_signal_chirp2}
	\end{align}	
	%\noindent\makebox[\linewidth]{\rule{18.3cm}{.4pt}}\vspace{-5mm}
\end{figure*}
\setcounter{equation}{14}
\subsubsection{Superimposed Chirp Waveforms}
If $\mathbf{x}_n(t)$ (given by \eqref{eq:Tx_chirp_Downlink_ecsi}) is transmitted from the AP, the signal ${y}_k^{[n]}(t)$ received at the $k^\text{th}$ user over its $n^\text{th}$ best subband is given by
\begin{align}
&y_k^{[n]}(t) \!=\! \sqrt{\Pt\eta_{[n],k}} {\mathbf{g}_{[n],k}^\top} {\pmb{\varphi}_{[n],k}^\ast} S_{[n],k}(t) \nonumber\\
&+ \sum_{\substack{j=1; j\neq k}}^K \sqrt{\Pt\eta_{[n],j}} {\mathbf{g}_{[n],k}^\top} {\pmb{\varphi}_{\{[n],k\},j}^\ast} S_{\{[n],k\},j}(t) {\mathds{1}_{\{[n],k\} \in \Xi_j}}  \nonumber\\
&+ w^A_{[n],k} (t),\label{eq:Rx_chirp_sensork_DLEBF2}
\end{align}
where $\mathbf{g}_{[n],k}^\top \in \mathbb{C}^{1\times M}$, $ \widehat{\mathbf{g}}_{[n],k} \in \mathbb{C}^{M\times 1}$ and $w^A_{[n],k} (t)$ denote the true channel vector, the estimated channel vector, and the antenna noise corresponding to the $k^\text{th}$ user over its $n^\text{th}$ best subband, respectively. The first term in \eqref{eq:Rx_chirp_sensork_DLEBF2} corresponds to the desired signal for the $k^\text{th}$ user over its $n^\text{th}$ best subband. The second term denotes the interference arising due to signals intended for other users over the $n^\text{th}$ best subband of the $k^\text{th}$ user. Thereby, any user $j \neq k$ will contribute to the interference term in $y_k^{[n]}(t)$, if the index of the $n^\text{th}$ best subband of the $k^\text{th}$ user belongs to $\Xi_j$.

\setcounter{equation}{24}
\begin{figure*}[t]\vspace{-4mm}
	\begin{equation} 
	p_k = \eta^2_{{[n],k}} \bigg[\frac{3}{2}\xi + 2 \binom{\xi}{2}\bigg] \sum_{n=1}^{\Nt}\Big(\gamma_k \varOmega^{[n]}_{k,2} +  \big(6\varOmega^{[n]}_{k,1} + \varUpsilon^{[n]}_k \big) \varUpsilon^{[n]}_k \Big) + \xi^2 \sum_{n \neq u}^{\Nt} \eta_{[n],k} \eta_{[u],k} \big(\varOmega^{[n]}_{k,1} + \varUpsilon^{[n]}_k \big) \big(\varOmega^{[u]}_{k,1} + \varUpsilon^{[u]}_k \big).\label{eq:harveat_energy_chirp_Qck41}
	\end{equation} 
	\begin{equation} 
	q_k = \beta^2_k \Bigg[\sum_{n=1}^{\Nt} \Bigg(\bigg[\frac{3}{2}\xi + 2 \binom{\xi}{2}\bigg]\sum_{\substack{j=1; j\neq k}}^K \eta^2_{[n],j} + \xi^2\sum_{\substack{j\neq p; j\neq k;p\neq k}}^K \eta_{[n],j}~\eta_{[n],p} \!\Bigg) + \xi^2 \sum_{n\neq u}^{\Nt} \Bigg(\sum_{\substack{j=1; j\neq k}}^K \eta_{[n],j} \sum_{\substack{j=1; j\neq k}}^K \eta_{[u],j} \!\Bigg) \Bigg].\label{eq:harveat_energy_chirp_Qck42} 
	\end{equation} %\vspace{-1mm}
	\noindent\makebox[\linewidth]{\rule{18.3cm}{.4pt}}\vspace{-5mm}
\end{figure*}
\setcounter{equation}{19}
A user receives a signal over the entire available bandwidth consisting of $N$ subbands and the received signal by the $k^\text{th}$ user $y_k(t) = \sum_{n=1}^N y_k^{[n]}(t) $ is given by \eqref{eq:total_received_signal_chirp}. As for each user, the signal is transmitted from the AP over its best $\Nt$ subbands independently, thereby, not all users will contribute to the power of the interference in the second term of \eqref{eq:total_received_signal_chirp} over all the $N$ subbands. Based on the set $\Xi_j$ of the best $\Nt$ subband entries corresponding to the $j^\text{th}$ user, we re-arrange the second term of \eqref{eq:total_received_signal_chirp} and obtain \eqref{eq:total_received_signal_chirp2}. With an input voltage proportional to $y_k(t)$ in \eqref{eq:total_received_signal_chirp2}, based on the nonlinearity of the diode, the output current at $k^\text{th}$ user is approximated as \cite{Besma,Clerckx1}
\setcounter{equation}{17}
\begin{align}
i_{\text{d}_k}(t) &= \sum_{m=1}^\infty \varepsilon_m (y_k(t))^m \approx \sum_{m=1}^4 \varepsilon_m (y_k(t))^m,
\label{eq:diode_current1}
\end{align}
where $\varepsilon_m=\frac{I_s}{m!} \left(\frac{1}{\vartheta V_T} \right)^m$, ${I_s}$ is the saturation current, $\vartheta$ is the emission coefficient, $V_T$ is the thermal voltage of the diode~\cite{Besma}. Note that, a linear energy harvesting model~\cite{Zhou2,Zhao,Lu} considers a second-order truncation of the expression in~\eqref{eq:diode_current1}, thereby neglecting higher order terms, mainly to maintain analytical simplicity rather than accuracy. However, the gain from the waveform design can not be observed by considering only the second-order term. Incorporating the higher order terms (i.e., $4^\text{th}$, $6^\text{th}$) brings out the nonlinear behavior of the diode, and contributes to the HE. Therefore, the infinite sum in~\eqref{eq:diode_current1} is truncated up to the fourth-order to maintain mathematical tractability, while preserving the nonlinearity behavior of the rectifier. This leads to an improvement in HE as compared to conventional linear energy harvesting techniques.

We consider that the energy to be stored in the battery is proportional to $\left[i_{\text{d}_k}(t) \right]_{\text{LPF}}$ where $\left[\cdot\right]_{\text{LPF}}$ indicates the diplexer LPF process that removes the high-frequency components starting at $f_n$ and above in $i_{\text{d}_k}(t)$ \cite{Besma}. Now, with a conversion efficiency of $\psi$, $0 < \psi \leq 1$, the equivalent average energy harvested by the $k^\text{th}$ user over $N$ subbands and over symbol duration $T$ is given by 
\begin{align}
Q_k &\approx \psi \E \bigg\{\int_0^T \bigg[\sum_{m=1}^{4} \varepsilon_m (y_k(t))^m \bigg]_{\text{LPF}} + w^D_k(t) dt\bigg\}  \nonumber\\	
&= \psi \E \bigg\{\int_0^T \bigg[\sum_{m=1}^4 \varepsilon_m (y_k(t))^m \bigg]_{\text{LPF}} dt\bigg\},
\label{eq:harveat_energy_chirp_sensork_ecsi}
\end{align}
where the expectation is with respect to the small-scale fading and $w^D_k(t) = \sum_{n=1}^N w^D_{[n],k} (t)$ with $w^{D}_{[n],k} (t) \sim \mathcal{CN}(0,\sigma^2)$ is the noise from the diode and diplexer over the $n^\text{th}$ subband. Furthermore, without loss of generality, we set $\psi=1$. Accordingly, the average HE is stated as follows.

\setcounter{equation}{26}
\begin{figure*}[t]\vspace{-5mm}
	\begin{equation}
		y^\prime_k(t) = \sum_{n=1}^N\Bigg(\sqrt{\Pt \eta_{[n],k}}  \mathbf{g}_{[n],k}^\top {\pmb{\varphi}_{[n],k}^\ast} S^{\prime}_{{[n],k}} (t) 
		+ \sum_{\substack{j=1; j\neq k}}^K \sqrt{\Pt\eta_{{[n],j}} }  \mathbf{g}_{[n],k}^\top {\pmb{\varphi}_{\{[n],k\},j}^\ast} S^{\prime}_{{\{[n],k\},j}}(t)\Bigg) + \sum_{n=1}^N w^A_{[n],k} (t).\label{eq:total_received_signal_ofdm}
	\end{equation}%\vspace{-3mm}
	\noindent\makebox[\linewidth]{\rule{18.3cm}{.4pt}}\vspace{-5mm}
\end{figure*}
\setcounter{equation}{19}

\begin{theorem}
The average HE $Q_k$ at the $k^{\text{th}}$ user based on transmission of superimposed chirp waveforms over $\Nt$ best subbands and over symbol duration $T$ under imperfect CSI and diode-nonlinearities is given by
\begin{align}
Q_k =&~ \varepsilon_2 \Pt \xi T \sum_{n=1}^{\Nt} \bigg[\eta_{[n],k} \big(\varOmega^{[n]}_{k,1} + \varUpsilon^{[n]}_k \big) + \beta_k \sum_{j=1; j\neq k}^K \eta_{[n],j} \bigg] \nonumber\\
& + 3\varepsilon_4 \Pt^2 T \Big[p_k + q_k + 2 r_k\Big],
\label{eq:thrm1_harveat_energy_chirp_sensork}
\end{align}
where
\begin{equation}
\varUpsilon^{[n]}_k = \frac{(\beta_k - \gamma_k) N!}{(n-1)!(N-n)!}\sum_{l= 0}^{N-n}\frac{(-1)^{N-n+l}}{(N-l)^2} \binom{N-n}{l},\label{eq:upsilon}
\end{equation}
\vspace{-2mm}
\begin{align}
\varOmega^{[n]}_{k,\nu} = \frac{\gamma_k N!}{(n-1)!(N-n)!}\sum_{l=0}^{n-1} &(-1)^l \binom{n-1}{l} \nonumber\\
\times &(\Gamma(M))^{n-N-l-1} \varLambda_{\nu}(l),\label{eq:omega}
\end{align}
\vspace{-6mm}
\begin{align}
\varLambda_{\nu}(l) &= M^{n-N-l} \Gamma(1+M(l+N-n+1)) \nonumber \\ &\quad\times F_A^{(l+N-n)} \big(\nu+M(l+N-n+1),M,\ldots,M; \nonumber\\
&\hspace{22mm} M+1,\ldots,M+1; -1,\ldots,-1 \big),\label{eq:thrm1_Omegaklambda}
\end{align}
with $\nu \in \{1,2\}$,
\begin{equation} %a3=5, 3/2
r_k = \xi^2 \beta_k \sum_{n=1}^{\Nt} \eta_{[n],k} \big(\varOmega^{[n]}_{k,1} + \varUpsilon^{[n]}_k \big) \sum_{u=1}^{\Nt} \sum_{j=1; j\neq k}^K \! \eta_{[u],j}, \label{eq:harveat_energy_chirp_Qck43} % (1-\eta_{[u],k})
\end{equation}
and $p_k$ and $q_k$, are given by \eqref{eq:harveat_energy_chirp_Qck41} and \eqref{eq:harveat_energy_chirp_Qck42}, respectively.\footnote{Note that $p_k$ indicates the contribution in HE due to the desired signal transmitted for the $k^\text{th}$ user, while $q_k$ indicates the contribution in HE arising due to the signals intended for the other users.}
\label{thm:avg_energy_chirp_SWEIT}
\end{theorem}

\begin{IEEEproof}
The proof is given in Appendix \ref{app:avg_energy_chirp_SWEIT}. 
\end{IEEEproof}

\begin{remark}
Analyzing \eqref{eq:thrm1_harveat_energy_chirp_sensork} leads to the following observations: 
	(i) The analysis above accounts for the order statistics of the subband channel estimates acquired through UL pilot signaling. 
	(ii) It brings out the dependence of the average HE $Q_k$ on the nonlinear characteristics of the diode. The first term of~\eqref{eq:thrm1_harveat_energy_chirp_sensork} presents the second-order term corresponding to the linear component of the energy harvesting model, whereas the second term (indicates the fourth-order term and corresponds to the nonlinear component of the energy harvesting model) is available only when a nonlinear energy harvesting model is considered. The second term is responsible for the nonlinear behavior of the diode~\cite{Besma,Clerckx1}, and indicates the nonlinearity gain of the diode. 
	(iii) Furthermore, the analysis also highlights the importance of the superimposed chirp in the form of $\xi$ as can be seen from~\eqref{eq:thrm1_harveat_energy_chirp_sensork}, where $\xi$ contributes as a scaling factor for HE, i.e., $Q_k$ improves as $\xi$ increases. Therefore, by increasing $\xi \leq N$, one can increase the average HE at the UT.		
	(iv) It indicates how the average HE scales as a function of the transmit power $\Pt$, the number of selected subbands $\Nt$, and the number of available bands $N$. 
	(v) $Q_k$ also depends on the power allocation strategy across subbands through $\eta_{[n],k}$ (e.g., power proportional to channel gain, equal power over all selected bands) and the number of users $K$. 	
	(vi) Analyzing \eqref{eq:thrm1_harveat_energy_chirp_sensork} indicates that the average HE is a function of the terms $\varOmega_{k}$ and $\varUpsilon_k$. Following the derivations in Appendix B, $\varOmega_{k}$ essentially captures the gain from channel ordering and beamforming gain while $\varUpsilon_k$ appears due to channel estimation error;  $\varUpsilon_k$ is much smaller in comparison to $\varOmega_{k}$. Thereby, both terms depend on the channel estimation through $\gamma_k$. Specifically, $\varOmega_{k}$ varies in proportion to $\gamma_k$. Thus, $Q_k$ depends on the variance $\gamma_k$ of the imperfect channel estimate which in turn depends on the pilot energy $E_{\text{p}}$. 
	(vii) Following that the beamforming precoder in \eqref{eq:beamform_Downlink_ecsi} is a function of the number of antennas $M$ at the AP, beamforming gain is also become a function (i.e., proportional) to $M$. While $Q_k$ is a function of $\varOmega_{k}$, $\varOmega_{k}$ is a function of the factor $M$ and is proportional to $M$. Thereby, through multiple antennas, the AP performs DL beamforming and contributes to the performance of $Q_k$.	
\end{remark}

\begin{remark}
When perfect CSI is available at the AP, $\beta_k - \gamma_k=0$. Thereby, replacing $\gamma_k$ with $\beta_k$ in \eqref{eq:thrm1_harveat_energy_chirp_sensork} we obtain $Q_k$ under perfect CSI scenario.
\end{remark}

\subsubsection{Fixed-frequency Waveform}\label{received_signal_ofdm}
Based on Remark \ref{downlink_beamform_ofdm} and by using \eqref{eq:Tx_ofdm_Downlink_ecsi} and \eqref{eq:total_received_signal_chirp}, the received signal $y^\prime_k(t)$ at the $k^\text{th}$ user over $N$ subbands and over symbol duration $T_\text{o}$ is given in \eqref{eq:total_received_signal_ofdm}. Now, replacing $y_k(t)$ with $y^\prime_k(t)$ in \eqref{eq:harveat_energy_chirp_sensork_ecsi}, we obtain the equivalent average HE $Q^\prime_k$ via transmission of fixed-frequency waveforms. 
Note that, with fixed-frequency, $N=\Nt$, i.e., $y^\prime_k(t)$ represents the signal with no subband selection. This is same as that of transmission of single chirp waveform $(\xi=1)$ over $N$ bands. Following the fact that the power allocation across subbands is in proportion to the channel gain, the subband ordering is preserved in \eqref{eq:total_received_signal_ofdm}. Therefore, using \eqref{eq:thrm1_harveat_energy_chirp_sensork}, the average HE $Q^\prime_k$ can be stated in closed-form as follows. 

\begin{corollary}
The average HE $Q^\prime_k$ at the $k^{\text{th}}$ user based on the transmission of fixed-frequency waveform over $N$ subbands and over symbol duration $T_\text{o}$ under imperfect CSI and diode-nonlinearities is given by
\setcounter{equation}{27}
\begin{align}
Q^\prime_k = \varepsilon_2 \Pt T_\text{o} &\sum_{n=1}^{N} \bigg[\eta_{[n],k} \big(\varOmega^{[n]}_{k,1} + \varUpsilon^{[n]}_k \big) + \beta_k \sum_{j=1; j\neq k}^K \eta_{[n],j} \bigg] \nonumber\\ 
&+ 3\varepsilon_4 \Pt^2 T_\text{o} \big[p^\prime_k + q^\prime_k + 2 r^\prime_k\big],
\label{eq:thrm1_harveat_energy_ofdm_sensork}
\end{align}
where $r^\prime_{k}$, $p^\prime_{k}$, and $q^\prime_{k}$ is obtained from \eqref{eq:harveat_energy_chirp_Qck43}, \eqref{eq:harveat_energy_chirp_Qck41}, and  \eqref{eq:harveat_energy_chirp_Qck42}, respectively, by replacing $\Nt$ with $N$. 
\label{thm:avg_energy_ofdm_SWEIT}
\end{corollary}%\vspace{-4mm}

\begin{IEEEproof}
The proof is based on order statistics and follows a similar methodology as in Appendix \ref{app:avg_energy_chirp_SWEIT} by setting $\xi=1$, and since $\binom{1}{2}=0$.
\end{IEEEproof}

The expression in \eqref{eq:thrm1_harveat_energy_ofdm_sensork} highlights the dependency of the average HE as a function of $\Pt$, $N$, $M$, and the power allocation strategy across subbands through $\eta_{[n],k}$. Moreover, comparing the expressions of $Q_k$ and $Q^\prime_k$ in \eqref{eq:thrm1_harveat_energy_chirp_sensork} and \eqref{eq:thrm1_harveat_energy_ofdm_sensork}, respectively, highlights the importance of the superimposed chirp that contributes in the form of multiplication factors in different terms of \eqref{eq:thrm1_harveat_energy_chirp_sensork} as compared to~\eqref{eq:thrm1_harveat_energy_ofdm_sensork}. This enables $Q_k$ to be greater than $Q^\prime_k$.

\begin{remark}
Given $Q_k$ and $Q^\prime_k$, the overall average HE at the $k^{\text{th}}$ user equals to $\frac{\tau_\text{DL}}{T} Q_k$ and $\frac{\tau_\text{DL}}{T_\text{o}} Q^\prime_k$ for superimposed chirps and fixed-frequency waveforms, respectively. 
\label{overall_harvested_energy_ofdm}
\end{remark}

\subsection{Average Received Information}\label{subsec:Analysis of DL Achievable Rate}

We analyze the average information received based on the transmission of superimposed chirp waveforms over selected subbands. For the ease of calculation we have dropped higher order terms and considered the first order term\footnote{It can be found from the expansion of the diode output that the signal components become smaller for higher-order terms. Further, the decomposition of higher-order terms introduces multiple noise terms due to expansion, and this leads to the effective SNR becoming smaller. Hence the contribution is not to be significant in the overall SNR.\vspace{-1mm}}. Furthermore, the signal energy is proportional to $\left[i_{\text{d}_k}(t) \right]_{\text{BPF}}$ where $\left[\cdot\right]_{\text{BPF}}$ indicates the RF diplexer BPF process that allows signal-frequency components starting at $f_n$ with bandwidth $B$, and removes other frequency components in $i_{\text{d}_k}(t)$ \cite{Besma}. For analytical tractability, we consider the superimposed chirp for a chirp-2 ($\xi=2$) configuration. Accordingly, from \eqref{eq:Rx_chirp_sensork_DLEBF2}, \eqref{eq:total_received_signal_chirp2} and \eqref{eq:diode_current1}, the equivalent signal at the input of the chirp demodulator corresponding to $k^{\text{th}}$ user over the $n^{\text{th}}$ best subband is 
\begin{align}
&y^{[n]}_k(t) = \varepsilon_1 \sqrt{\Pt\eta_{{[n],k}} }  \widehat{\mathbf{g}}_{[n],k}^\top {\pmb{\varphi}_{[n],k}^\ast} S_{{[n],k}}(t) \nonumber\\
&+ \varepsilon_1 \!\sum_{j=1; j\neq k}^K \sqrt{\Pt\eta_{{[n],j}} }  \widehat{\mathbf{g}}_{[n],k}^\top {\pmb{\varphi}_{\{[n],k\},j}^\ast}  S_{{\{[n],k\},j}}(t) {\mathds{1}_{\{[n],k\} \in \Xi_j}} \nonumber\\
&- \varepsilon_1 \sum_{j=1}^K \sqrt{\Pt\eta_{{[n],j}} } \widetilde{\mathbf{g}}_{[n],k}^\top {\pmb{\varphi}_{\{[n],k\},j}^\ast}  S_{{\{[n],k\},j}}(t) {\mathds{1}_{\{[n],k\} \in \Xi_j}} \nonumber\\
&+ w^{D}_{[n],k} (t),\label{eq:received_info_signal_chirp}
\end{align}
where the last line in \eqref{eq:received_info_signal_chirp} is related to the fact that ${\mathbf{g}}_{n,k}=\widehat{\mathbf{g}}_{n,k}-\widetilde{\mathbf{g}}_{n,k}$, and with the designed superimposed chirp waveforms, the chirps $s_{n,1}(t)$ and $s_{n,2}(t)$ in a subband for $\xi=2$ obtained from~\eqref{eq:signal_xi-chirp} are orthogonal to each other. 

Following, inside the chirp demodulator block, $y^{[n]}_k(t)$ is multiplied with the associated reference chirp $s_{[n],1}(t)$ followed by integration to obtain the demodulated signal $r_{[n],k,1}$ corresponding to the data symbol $d_{[n],k,1}$ to $k^{\text{th}}$ user over the $n^{\text{th}}$ best subband, and is given by
\begin{equation}
r_{[n],k,1} = \frac{1}{\sqrt{T}} \int_0^T y^{[n]}_k(t) s_{[n],1}(t) dt.
\label{eq:received_info_signal_chirp2}
\end{equation}

\begin{figure*}[t]\setcounter{equation}{31}
	{\begin{equation}
		\!\!\!\rho_{[n],k,1} \!=\! \E\Bigg\{\frac{\varepsilon^2_1 \Pt T \eta_{[n],k} \big|\widehat{\mathbf{g}}_{[n],k}^\top  {\pmb{\varphi}_{[n],k}^\ast} \big|^2 d^2_{[n],k,1}}
		{\!\varepsilon^2_1 \Pt T \!\bigg[\!\sum\limits_{\substack{j=1; j\neq k}}^K \!\eta_{[n],j} \!\big|\widehat{\mathbf{g}}_{[n],k}^\top {\pmb{\varphi}_{\{[n],k\},j}^\ast} \big|^2 \!d^2_{{\{[n],k\},1,j}} {\mathds{1}_{\{[n],k\} \in \Xi_j}} \!\! +\! \sum\limits_{j=1}^K \!\eta_{[n],j} \!\big|\widetilde{\mathbf{g}}_{[n],k}^\top {\pmb{\varphi}_{\{[n],k\},j}^\ast} \big|^2 \!d^2_{\{[n],k\},1,j} {\mathds{1}_{\{[n],k\} \in \Xi_j}} \!\bigg] \!\!+\! w^{D^2}_{[n],k}} \!\!
		\Bigg\}.\label{eq:avg_received_sinr_chirp}
		\end{equation}}\vspace{-1mm}\setcounter{equation}{34}
	\begin{equation}
	\rho_{[n],k} = \frac{\varepsilon^2_1 \Pt T \eta_{[n],k} \varOmega^{[n]}_{k,1} }{\varepsilon^2_1 \Pt T \Big[\E \Big\{ \sum_{j=1; j\neq k}^K \eta_{[n],j} \beta_k {\widetilde{\varUpsilon}^{[n]}_k} {\mathds{1}_{\{[n],k\} \in \Xi_j}} \Big\} + \eta_{[n],k} \varUpsilon^{[n]}_k \Big] + \sigma^2}.\label{eq:avg_rho_chirp_sensork}
	\end{equation}
	\noindent\makebox[\linewidth]{\rule{18.3cm}{.4pt}}\vspace{-5mm}
\end{figure*}\setcounter{equation}{30}
For each user, the signal is transmitted from the AP over its best $\Nt$ subbands belonging to the set $\Xi_k$, $k=1,2,\ldots,K$. Next, the output $r_{[n],k,1}$ is fed to a decision block, and with BPSK modulation~\cite[(III.A)]{Yang2}  the associated information, corresponding to the data symbol $d_{[n],k,1}$, transferred to the $k^{\text{th}}$ user over a symbol duration $T$ is
\begin{equation}
R_{k,1} \approx \sum_{n=1}^{\Nt} \sqrt{\frac{\rho_{[n],k,1}}{\rho_{[n],k,1} + 2}} \sum_{u=1}^\infty \frac{2(-1)^u}{\sqrt{1+\frac{2}{\rho_{[n],k,1}}} + 2u + 1},\label{eq:avg_received_info_chirp}
\end{equation}
where the average signal-to-interference-plus-noise ratio (SINR) $\rho_{[n],k,1}$ is given by \eqref{eq:avg_received_sinr_chirp} and\setcounter{equation}{32}
\begin{align}
w^D_{[n],k} = \frac{1}{\sqrt{T}} \int_0^T w^D_{[n],k} (t) s_{[n]}(t) dt.
\end{align}

In a similar manner, we can obtain $\rho_{[n],k,2}$ corresponding to the data symbol $d_{[n],k,2}$ for $k^{\text{th}}$ user by replacing $d_{[n],k,1}$ with $d_{[n],k,2}$ in \eqref{eq:received_info_signal_chirp2} (after multiplying $y^{[n]}_k(t)$ with the associated reference chirp $s_{[n],2}(t)$ followed by integration) and associated $R_{k,2} $ by using \eqref{eq:avg_received_info_chirp}. Simplifying the terms of \eqref{eq:avg_received_sinr_chirp} using \eqref{eq:harveat_energy_chirp_y22}, \eqref{eq:term_chirp_ecsi_17}, the information to the $k^{\text{th}}$ user can be stated as follows.
\begin{proposition}
The amount of information $R_k$ (in bits) that is conveyed via superimposed chirp based transmission to the $k^{\text{th}}$ user over $N$ subbands for a duration of $\tau_\text{DL}$ under estimated CSI is given by
\begin{equation}
R_k \approx \frac{\tau_\text{DL}}{T} \sum_{n=1}^{\Nt} 2\sqrt{\frac{\rho_{[n],k}}{\rho_{[n],k}+2}} \sum_{u=1}^\infty \frac{2(-1)^u}{\sqrt{1+\frac{2}{\rho_{[n],k}}}+ 2u+1},
\label{eq:prop_infor_chirp_sensork}
\end{equation}
where $\rho_{[n],k}$ is given in \eqref{eq:avg_rho_chirp_sensork} and the expectation is with respect to the indicator function that will take value $1$, if the index of the $n^\text{th}$ best subband of the $k^\text{th}$ user belongs to the set $\Xi_j$, else it would take value zero. Note, $\Xi_j$ holds the set of the best $\Nt$ subband entries corresponding to the $j^\text{th}$ user. Furthermore, ${\widetilde{\varUpsilon}^{[n]}_k}$ is given in~\eqref{eq:term_chirp_ecsi_18}.
\label{prop1:avg_infor_chirp_SWEIT}
\end{proposition}

Following Proposition \ref{prop1:avg_infor_chirp_SWEIT}, we present two limiting cases based on the value of indicator function $\mathds{1}_{\{[n],k\}} \in \Xi_j$. In the first case, we assume that the set of subband indices in $\Xi_j$ of the $j^\text{th}$ user, $j=1,2,\dots,K, j\neq k$, does not match to the set of subband indices in $\Xi_k$ of the $k^\text{th}$ user. This indicates $\mathds{1}_{\{[n],k\}} \in \Xi_j$ takes zero value $\forall j$, and accordingly $\rho_{[n],k}$ in \eqref{eq:avg_rho_chirp_sensork} simplifies to
\setcounter{equation}{35}
\begin{equation}
\rho_{[n],k} = \frac{\varepsilon^2_1 \Pt T \eta_{[n],k} \varOmega^{[n]}_{k,1} }{\varepsilon^2_1 \Pt T \eta_{[n],k} \varUpsilon^{[n]}_k + \sigma^2 },\label{eq:avg_rho_chirp_sensork_upper}
\end{equation}
which indicates improvement in SINR. Substituting \eqref{eq:avg_rho_chirp_sensork_upper} in \eqref{eq:prop_infor_chirp_sensork} gives the upper-bound on the received information to the $k^\text{th}$ user. On the other hand, in the second case, consider that $\mathds{1}_{\{[n],k\}} \in \Xi_j$ takes 1 $\forall j$, i.e., the set of subband indices in $\Xi_j$ $\forall j$ matches to the subband indices in $\Xi_k$. Accordingly $\rho_{[n],k}$ in \eqref{eq:avg_rho_chirp_sensork} simplifies to
\begin{equation}
\rho_{[n],k} = \frac{\varepsilon^2_1 \Pt T \eta_{[n],k} \varOmega^{[n]}_{k,1} }{\varepsilon^2_1 \Pt T \Big(\sum_{j=1; j\neq k}^K \eta_{[n],j} \beta_k {\widetilde{\varUpsilon}^{[n]}_k} + \eta_{[n],k} \varUpsilon^{[n]}_k \Big) + \sigma^2}.
\label{eq:avg_rho_chirp_sensork_lower}
\end{equation}
This indicates degradation in SINR due to increase in interference. Substituting \eqref{eq:avg_rho_chirp_sensork_lower} in \eqref{eq:prop_infor_chirp_sensork} gives the lower-bound on the received information to the $k^\text{th}$ user.

For the case of fixed-frequency waveform, we consider a similar integrated receiver structure as that shown in Fig. \ref{fig:integrated_rx_sweit} with the chirp demodulator replaced by a demodulator for fixed-frequency signals. Further, with fixed-frequency, signals are transmitted over all available $N$ subbands without any selection. Following a similar technique as used above for chirp waveform, based on Section \ref{received_signal_ofdm} and using \eqref{eq:total_received_signal_ofdm}, \eqref{eq:received_info_signal_chirp}, the information to the $k^{\text{th}}$ user can be stated as follows.\label{demod_signal_ofdm}

\begin{proposition}
The amount of information $R^\prime_k$ (in bits) that is conveyed via fixed-frequency waveform based transmission to the $k^{\text{th}}$ sensor over $N$ subbands for a duration of $\tau_\text{DL}$ seconds under estimated CSI is given by
\begin{equation}
R^\prime_k \approx\! \frac{\tau_\text{DL}}{T_\text{o}} \sum_{n=1}^N \sqrt{\frac{\rho^\prime_{[n],k}}{\rho^\prime_{[n],k}+2}} \sum_{u=1}^\infty \frac{2(-1)^u}{\sqrt{1+\frac{2}{\rho^\prime_{[n],k}}}+ 2u+1},\!	\label{eq:prop_infor_ofdm_sensork}
\end{equation}
where
\begin{equation}
\rho^\prime_{[n],k}	= \frac{\varepsilon^2_1 \Pt T_\text{o} \eta_{[n],k} \varOmega^{[n]}_{k,1} }{\varepsilon^2_1 \Pt T_\text{o} \Big[\sum_{j=1}^K \eta_{[n],j} \beta_k {\widetilde{\varUpsilon}^{[n]}_k} - \eta_{[n],k} \gamma_k {\widetilde{\varUpsilon}^{[n]}_k} \Big] + \sigma^2}.\label{eq:avg_rho_ofdm_sensork}
\end{equation}\label{prop2:avg_infor_ofdm_SWEIT}\vspace{-4mm}
\end{proposition}

\begin{remark}
	Analyzing the expressions in \eqref{eq:prop_infor_chirp_sensork} and \eqref{eq:prop_infor_ofdm_sensork} with $T=2T_\text{o}$, we observe that the received information is a function of the transmitted signal bandwidth. Specifically, it depends on the number of selected subbands $\Nt$ for the superimposed chirps, and the subbands $N$ for fixed-frequency waveforms. The expressions also indicates a linear improvement in received information with the increase of the number of bands. On the other hand, the analysis of average HE with superimposed chirp, in \eqref{eq:thrm1_harveat_energy_chirp_sensork}, indicates that with the increase of $\xi$ (decrease $\Nt$), the HE increases. Following these remarks, we observe that although superimposed chirp achieves a superior HE performance over fixed-frequency waveform, it attains a lower information rate due to not operating over all $N$ bands. This means as we increase $\xi$, it allows us to harvest more energy while reduces the achievable information amount at the receiver, and vice-versa. Further, $\xi=1$, $N=\Nt$, basically refers to the transmission of a single chirp waveform over $N$ bands that leads to the similar HE and received information performance as that of fixed-frequency waveform. Therefore, by varying $\xi$, superimposed chirp allows to maintain a trade-off between the HE and received information at the UT. The selection of $\xi$ and associated $\Nt$ thus becomes an application specific parameter.  					   
\end{remark}

\vspace{-2mm}
\section{Numerical Results}
We present numerical results to illustrate the potential of using superimposed chirps for SWIPT operation. The large-scale fading coefficient for the $k^{\text{th}}$ user is modeled as $\beta_k \!=\! 10^{-3}D_k^{-3}$ \cite{Yang}, where $D_k$ is the distance, in meters (m), between the AP and the $k^\text{th}$ user. Without loss of generality, we evaluate the system performances considering $\xi \!=\! 2$ with $\widetilde{N} \!=\! \frac{N}{2}$, $B_\text{total} \!=\! 3.2$ MHz, $N \!=\! 16$, and $B \!=\! 200$ KHz. Furthermore, we take $\Pt \!=\! 1$ W, $\sigma^2 \!=\! 10^{-20}$ J, and $\tau_\text{ch} \!=\! 190.5$ ms. 
The parameters for the nonlinear EH model are taken as $V_T = 25$ mV, $\vartheta \approx 1$, and $I_s = 0.6$ mA \cite{Besma}. In order to validate the mathematical analysis, we also present the corresponding Monte Carlo simulation results where we average over $10^5$ independent channel realizations. %Considering the receiver as a small sensor,

\begin{figure}[!t]\centering\vspace{-4mm}
	\includegraphics[width=0.8\linewidth]{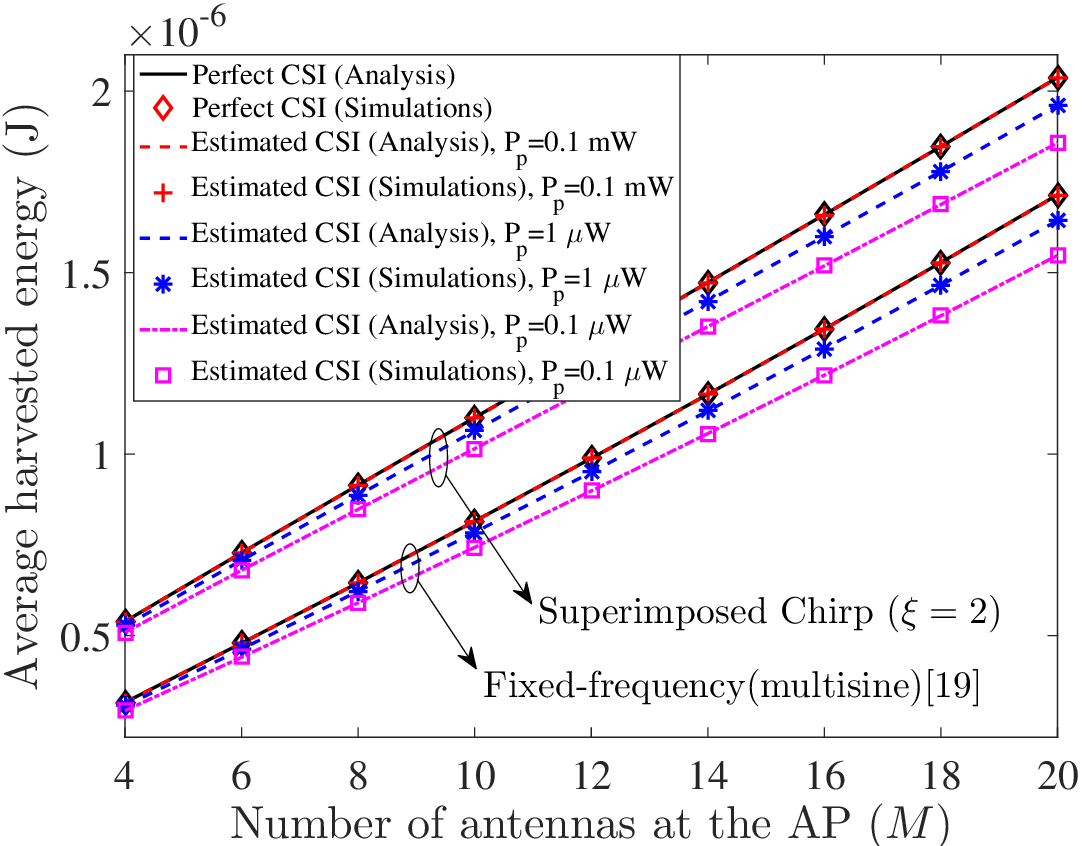}
	\caption{Effect of superimposed chirp waveform along with channel estimation error on average HE ($N=16$, $K = 1$, and $\xi = 2$).}\label{fig:chn_err_HE_chirp}\vspace{-4mm}
\end{figure}
Fig. \ref{fig:chn_err_HE_chirp} depicts the average HE as a function of $M$ with $N=16$, $\Nt = 8$ for single user placed at a distance of $D = 10$ m from the AP while performing power allocation in proportion to subband gain. The results indicate significant gains in HE with the employment of the proposed superimposed chirp waveforms in comparison to multisine waveforms~\cite{Clerckx1} consisting of a set of fixed-frequency cosine signals. For multisine waveforms~\cite{Clerckx1}, the power of the transmitted signal over a subband is considered to be proportional to the channel gain, while no subband selection is taken into account. Whereas our proposed transmission scheme performs power allocation in proportion to the channel gains over the set of selected subbands. Thereby, with superimposed chirps more energy is pushed over the set of subbands having relatively higher channel gains, thus exploiting frequency diversity in a much higher degree. For example, with $M = 12$, superimposed chirps lead to a gain of around $30$\% in average HE over fixed-frequency (multisine) waveforms. Furthermore, Fig. \ref{fig:chn_err_HE_chirp} illustrates the impact of pilot power on the average HE for different values of $P_\text{p}$. We observe that as the pilot power increases, the quality of channel estimation improves (i.e., channel estimation error decreases) and the average HE improves. Results also indicate that under imperfect CSI at a pilot power $P_\text{p} = 0.1$ mW, the HE is very close to that obtained under perfect CSI. On the other hand, by decreasing $P_\text{p}$ (from $0.1$ mW to $0.1 \mu$W), the HE performance decreases due to degradation in the quality of channel estimates. Note that the average HE increases with $M$ due to increase in the beamforming gain. For all cases, the analytical expressions match closely with Monte Carlo simulations, thereby verifying our proposed analytical framework.

\begin{figure}[!t]\centering\vspace{-4mm}
\includegraphics[width=0.80\linewidth]{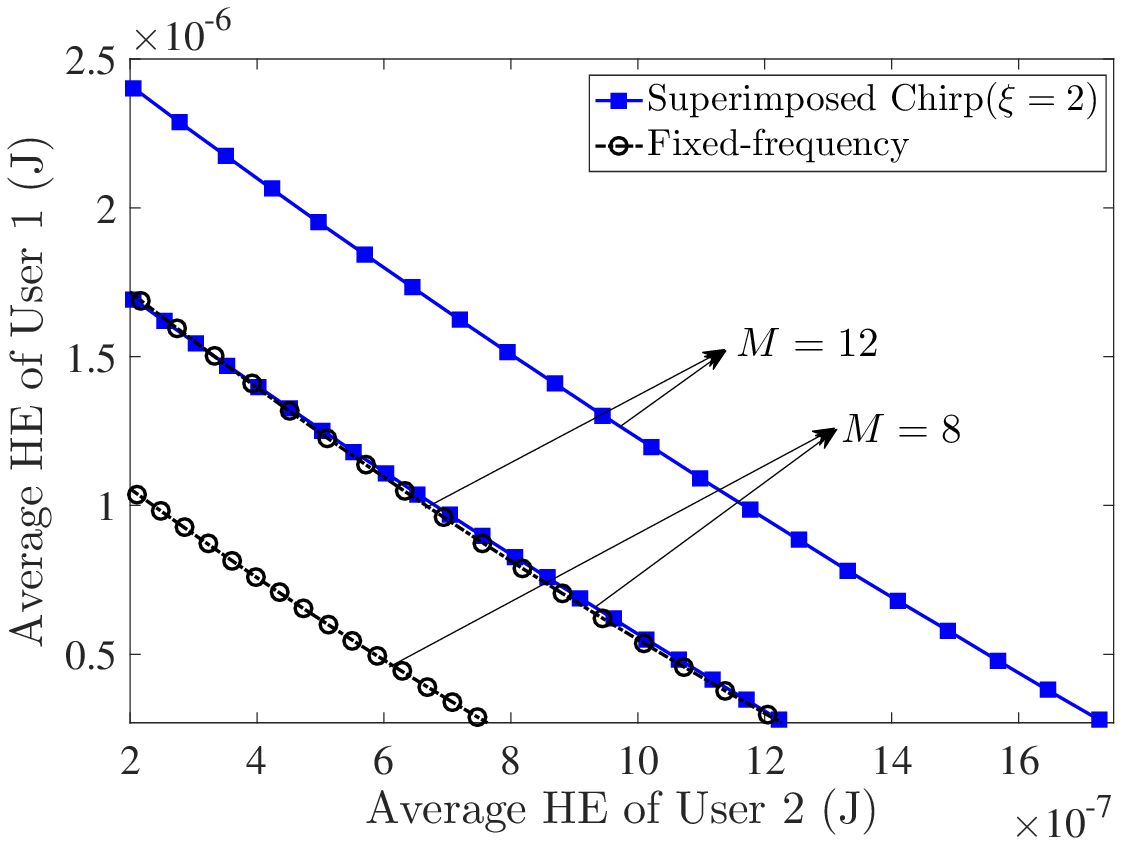}
\caption{Effect of power-allocation weights $\zeta_k$ on average HE region ($M=8,12$, $N=16$, and $K = 2$ with $D_1=8.2$ m, $D_2=9.1$ m).}\label{fig:weights_HE_chirp}\vspace{-4mm}
\end{figure}
Fig. \ref{fig:weights_HE_chirp} draws some useful insights into the average HE region by varying the power-allocation weights $\zeta_k$, $0\leq \zeta_k \leq 1$, in a two-user scenario as a function of $M$ where user 1 and user 2 are placed at distances of $8.2$ and $9.1$ m from the AP, respectively. Here $\zeta_k$ indicates the fraction of the average transmit power $\Pt$ that is allocated to $k^\text{th}$ user at the AP. According to transmission strategy, $\zeta_k$ is distributed over the selected subbands in proportion to their channel gains through the power control coefficients $\eta_{[n],k}$ such that $\zeta_k=\sum_{n=1}^{\Nt} \eta_{[n],k}$. Considering $P_{\text{tx}_k}$ is the power allocated to $k^\text{th}$ user, then we have $P_{\text{tx}_k} = \zeta_k \Pt = \Pt \sum_{n=1}^{\Nt} \eta_{[n],k}$. From Fig. \ref{fig:weights_HE_chirp}, we make the following observations$-$ firstly, the proposed superimposed chirps provide a wider HE region over fixed-frequency waveforms. Secondly, the region of HE increases with $M$ since both type of waveforms benefits from beamforming gain. Thirdly, the result demonstrates that the use of superimposed chirps achieve a similar HE region as that of fixed-frequency while employing significantly less number of antennas at the AP. For example, it requires $33\%$ less antennas to attain the same HE with superimposed chirps as compared to fixed-frequency waveforms. With less number of antennas, the operational energy requirement at the AP decreases, thereby, energy efficiency of the network for SWIPT increases by employing superimposed chirps over fixed-frequency waveforms. Alternatively, the result demonstrates that, superimposed chirp improves the minimum HE level in the network. For example, with $M=12$, when user 1 attains a HE of $1.5\times10^{-6}$ J, with fixed-frequency user 2 have a HE of $4\times10^{-7}$ J, whereas, with superimposed chirp user 2 achieves a HE of $8.1\times10^{-7}$ J which is more than double.

\begin{figure}[!t]\centering \vspace{-4mm}
	\subfigure[]{\includegraphics[width=0.49\linewidth]{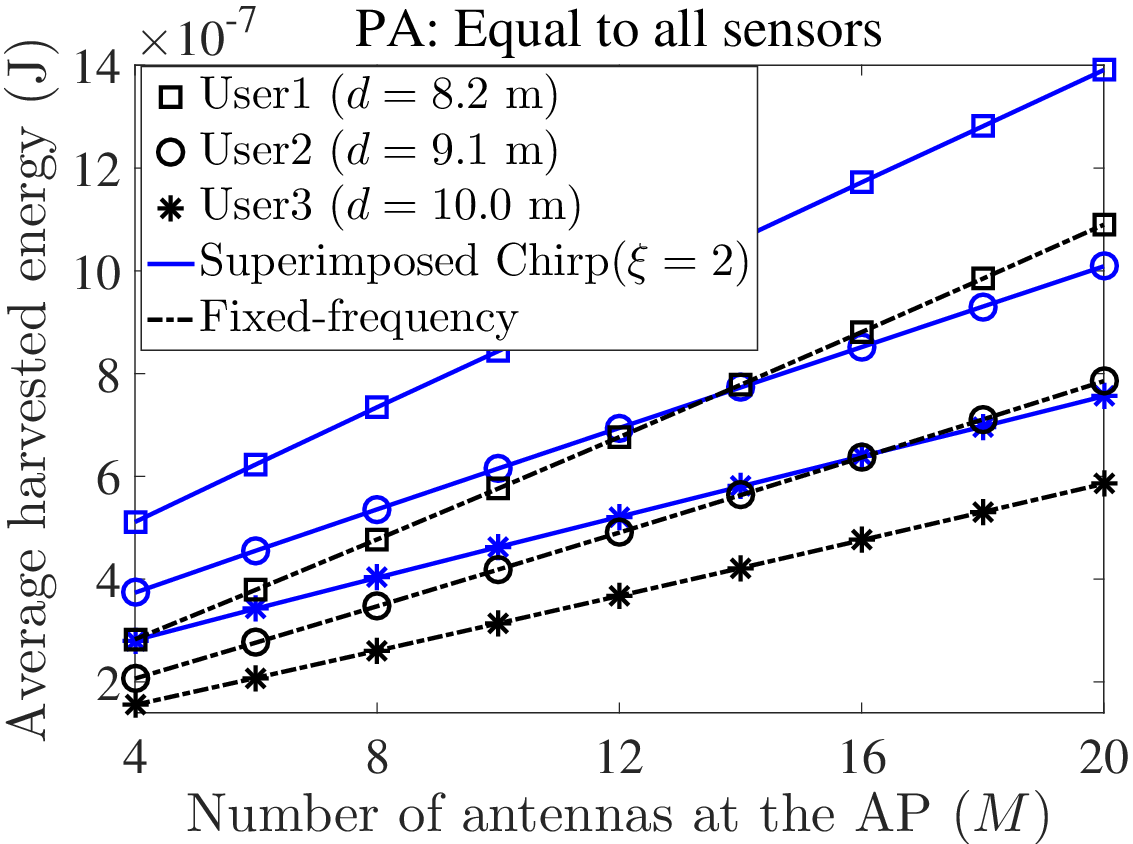}}\hfill
	\subfigure[]{\includegraphics[width=0.49\linewidth]{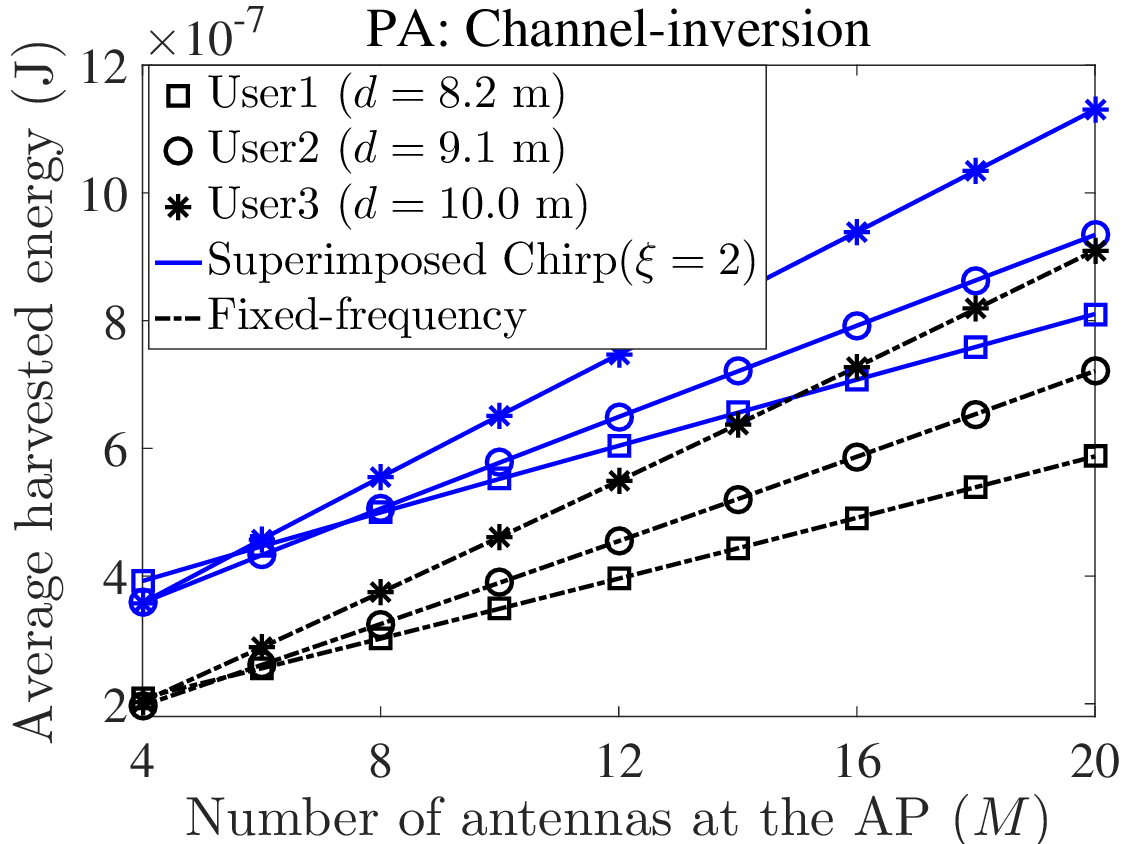}}
	\caption{Effect of waveforms on the average HE for different PA policies ($N \!=\! 16$, $K \!=\! 3$) (a) Equal PA. (b) Channel inversion PA.}\label{fig:powerAlloc_HE_chirp}\vspace{-1mm}
\end{figure}

\begin{figure}[!t]\centering
	\includegraphics[width=0.8\linewidth]{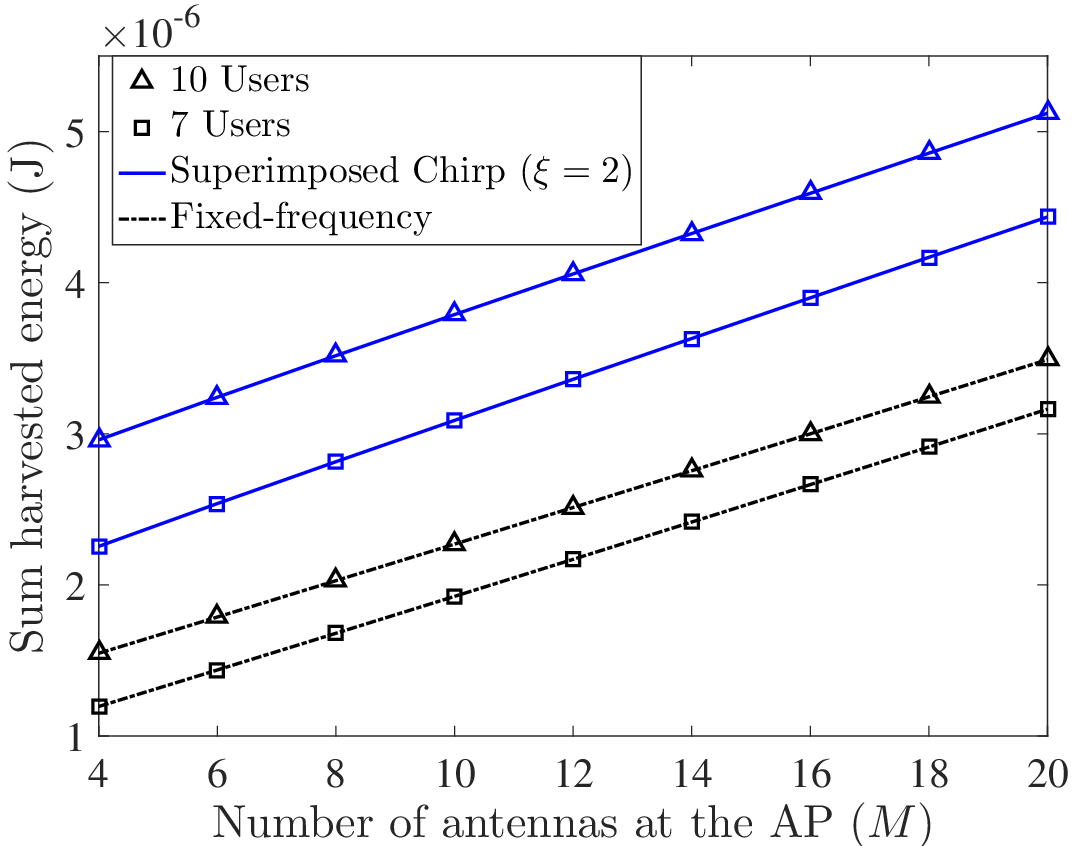}
	\caption{Effect of number of users on the sum HE ($N=16$, and $K = 7, 10$).}\label{fig:sum_HE_chirp}\vspace{-4mm}
\end{figure}
Fig. \ref{fig:powerAlloc_HE_chirp} presents the effect of waveforms on the average HE for different power allocation (PA) policies as a function of $M$, $N = 16$, and $K = 3$ users placed at a distance of $8.2$, $9.1$, and $10$ m, respectively from the AP. We consider two indicative PA schemes namely (i) equal PA, and (ii) channel inversion PA. With equal PA scheme, allocated power for each user is the same, i.e., $\zeta_k=\frac{1}{K}$. The associated average HE, shown in Fig. \ref{fig:powerAlloc_HE_chirp}(a), indicates that the user nearest to the AP harvests most energy since the distance-dependent path loss is minimum while the farthest user harvest the least. And for the same reason, although the HE increases with $M$, the difference in HE also increases. We also observe that the HE for the superimposed chirp outperforms the fixed-frequency waveform indicating that the proposed waveform successfully exploits the frequency diversity of the corresponding channel. 
On the other hand, with channel inversion PA, $\zeta_k$ is calculated as $\zeta_k=\frac{1}{\beta^2_k}/\frac{1}{\sum_{j=1}^K \beta^2_j}$ \cite{Yang} such that the farther the user from the AP, the more power is allocated for it to composite the path loss. Fig. \ref{fig:powerAlloc_HE_chirp}(b) plots the associated average HE at the users. We observe that for lower $M$, users placed at different distances harvest similar level of energy. However, at higher $M$, user placed at farther distance attain a higher HE due to the boost in beamforming gain. Note that, with channel inversion PA, the difference in HE among different users is much smaller over the equal PA scenario. Furthermore, with channel inversion PA also, superimposed chirp outperforms the fixed-frequency in terms of HE indicating the effectiveness of the proposed waveform design and transmission strategy. 

Fig.~\ref{fig:sum_HE_chirp} depicts the sum HE as a function of $M$, while considering a set of $7$ users and $10$ users, respectively. Users are placed uniformly at a distance of $7.6-10.0$ m from the AP. An equal gain PA is considered for allocating power among the users. With superimposed chirp, the AP can serve each user over the set of the subbands having higher gains. Therefore, by taken into account the frequency diversity gain attained by each user, superimposed chirp is able to provide a higher diversity gain as compared to fixed-frequency (multisine) waveforms. This enables to achieve a significantly higher sum HE using superimposed chirps over fixed-frequency waveforms. For example, considering $M = 12$ and $7$ users, superimposed chirp leads to a gain of around $54\%$ in HE sum over fixed-frequency waveform. On there other hand, as the number of users increases, the radiated power from the AP is captured by more UTs. It leads to an increase in HE sum, as seen from Fig.~\ref{fig:sum_HE_chirp}. For example, considering $M = 12$, with superimposed chirp, we observe a $21\%$ increase in HE sum for $10$ users over $7$ users; and with fixed-frequency waveform, observe an increase of $16\%$ in HE sum for $10$ users over $7$ users.

\begin{figure}[!t]\centering \vspace{-4mm}
\includegraphics[width=0.8\linewidth]{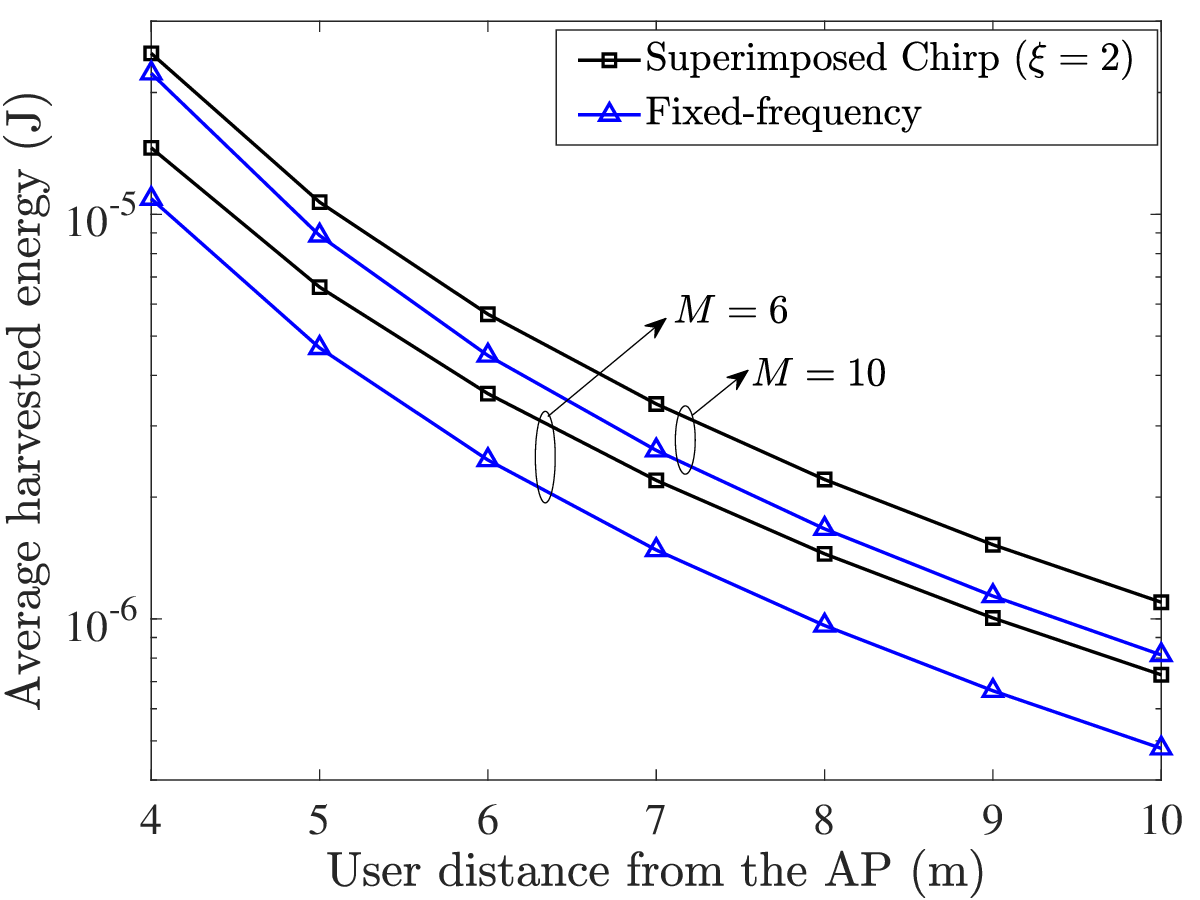} 
\caption{Effect of distance on average HE ($M=6,10$, $N=16$, and $K = 1$).}\label{fig:distance_HE_chirp}
\end{figure}
Fig. \ref{fig:distance_HE_chirp} plots the average HE as a function of the distance from the AP. For the superimposed chirp and the fixed-frequency waveforms for $N = 16$, we show results for $M = 6$ and $10$. We observe that chirp waveform boosts the operational range of SWIPT relative to fixed-frequency waveform. For example, with $M \!=\! 6$ antennas at the AP, SWIPT based on superimposed chirp can extend the range of HE by $1$ m over fixed-frequency waveform while maintaining the average HE at $10^{-6}$ J. Results also depict that the operation range of SWIPT can be extended by increasing the number of antennas at the AP while maintaining the HE at the same target level. 

\begin{figure}[!t]\centering\vspace{-4mm}
	\includegraphics[width=0.80\linewidth]{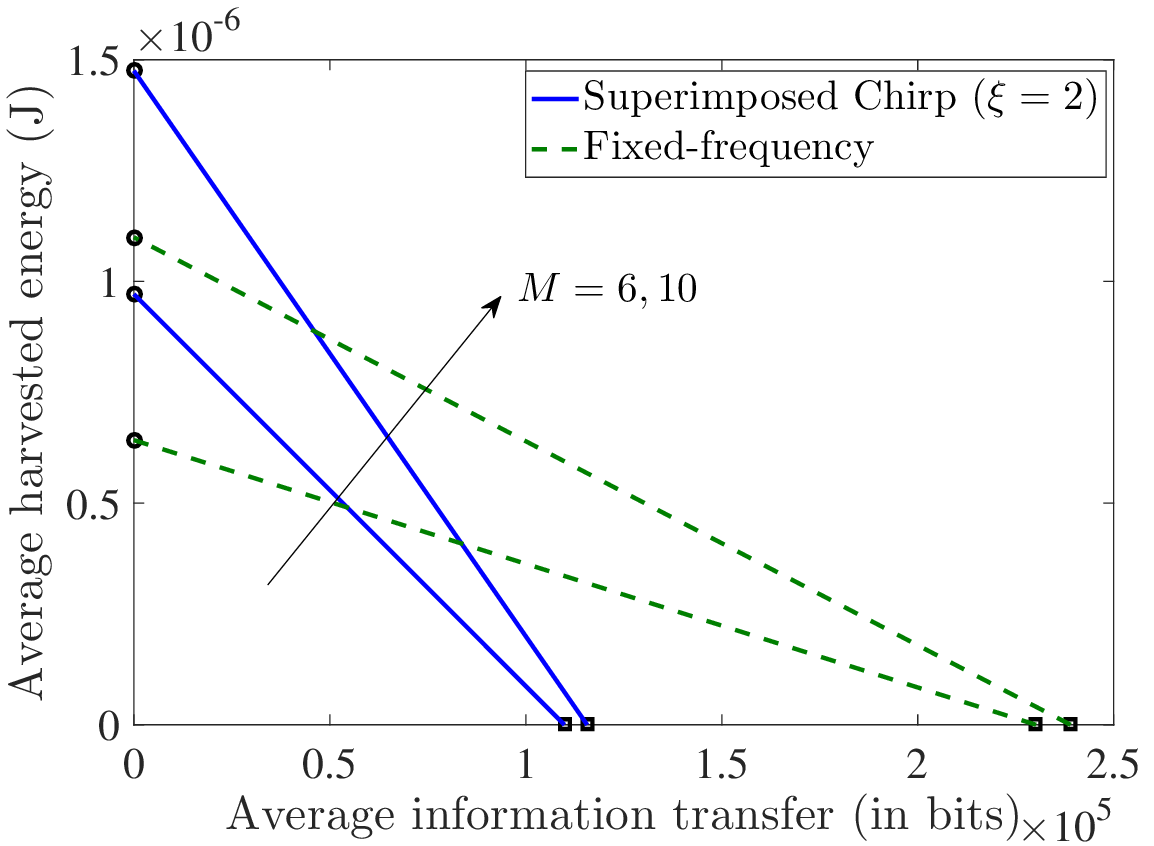}
	\caption{Energy-information transfer region ($M=6,10$, $N=16$ and $K = 1$).}
	\label{fig:avg_HE_info_chirp}\vspace{-4mm}
\end{figure}
Fig. \ref{fig:avg_HE_info_chirp} illustrates the energy-information transfer region for DL SWIPT via transmission of superimposed chirp and fixed-frequency waveform as a function of $M=6,10$ with $N=16$ and $K = 1$ user placed at a distance of $9.1$ m from the AP. From the analysis of the amount of information in \eqref{eq:prop_infor_chirp_sensork} and \eqref{eq:prop_infor_ofdm_sensork} indicates that the improvement in received information varies linearly with the number of subbands $N$ rather than the SINR level. Following the fact that the superimposed chirp operates over a set of subbands $\Nt<N$ with higher channel gains, it achieves a superior HE level over fixed-frequency, but attains a lower information rate compared to fixed-frequency waveform (due to not operating over all $N$ bands). Furthermore, the scenario $\xi = 1$, $N = \Nt$, basically refers to the transmission of a single chirp waveform over $N$ bands that leads to a similar HE and received information performance as that of fixed-frequency waveform. Thus, by varying $\xi$, superimposed chirp allows to attain a trade-off in terms of HE and received information performance. In addition, since the majority of the SWIPT applications would require a low and limited information traffic, using superimposed chirp can fulfill the required information rate while accomplishing the more challenging demand of achieving a higher HE level and/or improving the effective operational range. 
 
\begin{figure}[!t]\centering\vspace{-4mm}
	\includegraphics[width=0.8\linewidth]{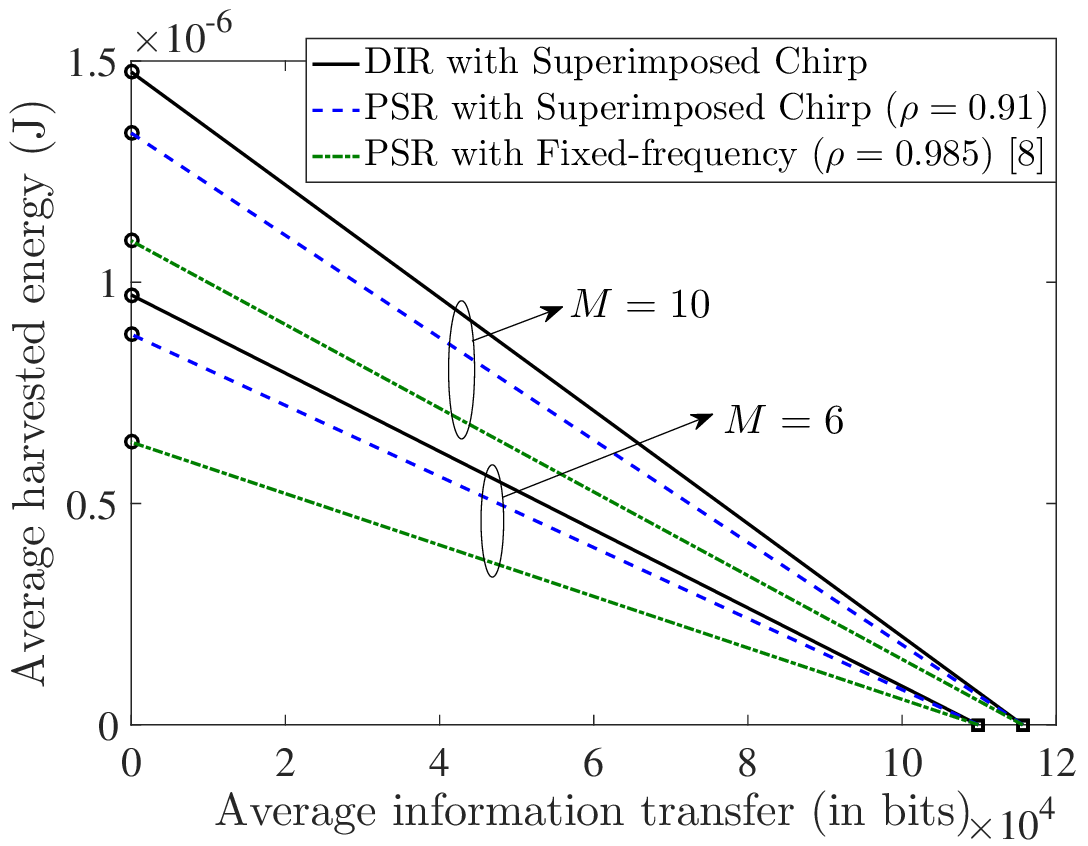}
	\caption{Performance comparison of the RF diplexer-based integrated receiver (DIR) with superimposed chirp vs. the power splitting receiver (PSR) with superimposed chirp and fixed-frequency waveform ($N=16$, $K = 1$).} 
	\label{fig:Comp_avg_HE_info_chirp}\vspace{-4mm}
\end{figure}

Fig. \ref{fig:Comp_avg_HE_info_chirp} compares the energy-information transfer performance of the DIR with superimposed chirp against a power splitting receiver (PSR) \cite{Wang,Zhou} with fixed-frequency as well as superimposed chirp waveforms under different antenna configurations with $N=16$ and $K=1$ user. A PSR is a widely used and accepted model in SWIPT literature~\cite{Zhou,Wang,Zhou2} where the received signal is split into two streams with the power ratio of $\rho$ and $1-\rho$, to be used for energy harvesting and information decoding, respectively, where $0\leq \rho \leq 1$. BPSK modulation is considered for both type of receivers. In contrary, DIR allows the same received signal to be used for energy harvesting and information decoding, and therefore for a certain threshold information rate, achieving a higher HE performance which leads to a broader energy-information transfer region over PSR. The result indicates that PSR with fixed-frequency waveform devote a very small faction of the signal power ($\rho=0.985$) to attain the threshold information rate as signals are transmitted over all subbands, whereas, for superimposed chirp it needs to devote higher signal power with a $\rho$ of $0.91$ to maintain the same information level. However, the frequency diversity gain attained by the superimposed chirp allows to meet a higher HE level as compared to the fixed-frequency scenario. Thereby, PSR with superimposed chirp outperforms PSR with fixed-frequency. Lastly, the result shows with increasing $M$, we achieve a wider energy-information transfer region while maintaining the similar receiver characteristics. 

\section{Conclusion}
This paper has studied the gains of using superimposed chirp over fixed-frequency waveforms for DL-SWIPT operation from a multi-antenna AP to a group of DIR-based users. Specifically, for SWIPT applications, we presented a general design of the superpimposed chirp waveforms. Furthermore, for improving the HE, incorporating the properties of chirp, a subband selection based transmission scheme is presented to attain the benefits of diode nonlinearity and frequency diversity. We derived new closed-form analytical expressions for average HE, in a multiuser network, based on the transmission of superposed chirp as well as fixed-frequency waveforms. The analysis takes into account the nonlinearity of the harvester and exploits the order statistics of the subband channel estimations. We also derived the DL average information rate achieved at the user. 
Through the analysis and simulations, for DL-SWIPT, we proved the following key points$-$ superimposed chirp provides considerable improvement in the average HE at a user, the minimum HE level in a multiuser network, and extends the operational range of energy transfer over fixed-frequency while maintaining a certain HE level. The integration of DIR with superimposed chirp enables the user to attain a higher HE as compared to conventional PS receiver based on chirp or fixed-frequency while keeping the information transfer at the same level. This would allow the user to maintain the SWIPT operation over a longer period. Therefore, as a whole the proposed transmission strategy combined with superimposed chirp successfully brings out the combined benefits of diode nonlinearity and frequency diversity gains.

\vspace{-2mm}
\appendix 
\subsection{Proof of $\sum_{k=1}^K \sum_{n=1}^N \eta_{n,k} \leq 1/\xi$}
\label{app:Tx_power_constraint_chirp_SWEIT}
A superposition $S_m(t)$ of the set of chirp waveforms as described in (\ref{eq:signal_xi-chirp}) over subband $m$ is mutually orthogonal to the superposition $S_n(t)$ of the set of chirp waveforms over a distinct subband $n \neq m$. In other words, they satisfy the following relation
\begin{equation}
\vspace{-0.5mm}
\int_0^T S_m(t) S_n(t) dt = 0,\label{eq:xi-chirp_cross-corr}
\end{equation}
where $S_i(t) = s_{i,1}(t) + s_{i,2}(t) + \dots + s_{i,\xi}(t)$, $i \in \{1,2,\dots,N\}$. Moreover, for $m=n$, we have
\begin{equation}
\int_0^T S_m(t)^2 dt = \int_0^T \big(s_{m,1}(t)^2 + s_{m,2}(t)^2 + \dots + s_{m,\xi}(t)^2 \big) dt.\label{eq:xi-chirp_uncorrelated}
\end{equation}
We can now prove $\sum_{k=1}^K \sum_{n=1}^N \eta_{n,k} \leq 1/\xi$, which ensures that the average transmit power constraint is satisfied at the AP. By using \eqref{eq:Tx_chirp_Downlink_ecsi} and the above two relations, the average transmit power of the AP is given by
\begin{align}
\vspace{-0.5mm}
\sum_{n=1}^N \E\left\{\left\|\mathbf{x}_n(t) \right\|^2 \right\} &= \Pt \sum_{k=1}^K \sum_{n=1}^{\Nt} \E\bigg\{\eta_{[n],k} \big\|{\pmb{\varphi}_{[n],k}^\ast} \big\|^2 \nonumber\\
&\times \frac{2}{T} \int_0^T \sum_{l=1}^\xi d^2_{[n],k,l} ~ s_{[n],l}(t)^2 dt \bigg\}.	\label{eq:Tx_power_cnstrain_chirp_proof4} 
\end{align}
Note that, for each user, the signal is transmitted over its best $\Nt$ subbands. Thus the summation is evaluated according to the indices corresponding to the $\Nt$ best subbands. Moreover, since each chirp waveform maintains average power of unity, \eqref{eq:Tx_power_cnstrain_chirp_proof4} simplifies to
\begin{align}
\vspace{-0.5mm}
\sum_{n=1}^N \E\left\{\left\|\mathbf{x}_n(t) \right\|^2\right\} &= \Pt \sum_{k=1}^K \sum_{n=1}^{\Nt} \xi \eta_{[n],k} \E\Bigg\{\frac{\big\|{\widehat{\mathbf{g}}_{[n],k}}\big\|^2} {\big\|{\widehat{\mathbf{g}}_{[n],k}}\big\|^2} \Bigg\} \nonumber\\
&=\Pt \sum_{k=1}^K \sum_{n=1}^{\Nt} \xi\eta_{[n],k} \leq \Pt.\label{eq:Tx_power_cnstrain_chirp_proof5}
\end{align}
As the total transmit power is constrained to $\Pt$, from \eqref{eq:Tx_power_cnstrain_chirp_proof5} we have $\sum_{k=1}^K \sum_{n=1}^N \eta_{n,k} \leq 1/\xi$.

\vspace{-1.5mm}
\subsection{Proof of Theorem \ref{thm:avg_energy_chirp_SWEIT}}
\label{app:avg_energy_chirp_SWEIT}
In order to evaluate the average energy $Q_k$ harvested by the $k^\text{th}$ user, we need to compute $(y_k(t))^m$ for $1 \leq m \leq 4$, where ${y}_k(t)$ is given by \eqref{eq:total_received_signal_chirp2}. For simplicity, $w^A_{[n],k} (t)$ is considered to be negligible and not included in the evaluation of the average HE \cite{Sotiris}.

For $m=1$, we have $\varepsilon_1\E \{\int_0^T [y_k(t)]_{\text{LPF}} dt \}=0$ following the fact that waveforms at frequency $f_n(t)$ are removed. In what follows, to facilitate our analysis, we define
\begin{align}
\vspace{-0.5mm}
v_k(t) = \sum_{n=1}^{\Nt} \sqrt{\Pt \eta_{[n],k}} \mathbf{g}_{[n],k}^\top {\pmb{\varphi}_{[n],k}^\ast} S_{[n],k}(t), \nonumber\\
v_j(t) = \sum_{j=1; j\neq k}^K \sum_{n=1}^{\Nt} \sqrt{\Pt\eta_{[n],j}}  \mathbf{g}_{\{[n],j\},k}^\top {\pmb{\varphi}_{[n],j}^\ast}  S_{[n],j}(t),
\end{align}
with $k\neq j$. Then, for $m=2$, expanding $y_k(t)^2$ gives
\begin{equation}
\vspace{-0.5mm}
y_k(t)^2 = v_k(t)^2 + v_j(t)^2 + 2 v_k(t) v_j(t).\label{eq:harveat_energy_chirp_y21}
\end{equation}
Following the fact that $S_{n,k}(t)$ is a periodic and even function, the expectation of the time-average for the $3^\text{rd}$ term of \eqref{eq:harveat_energy_chirp_y21} is zero.

Therefore, $\mathcal{E}_1 = \varepsilon_2\E \big\{\int_0^T [{y}_k(t)^2]_{\text{LPF}} dt \big\}$ is evaluated as
\begin{align}
&\mathcal{E}_1 = \varepsilon_2 \Pt \sum_{n=1}^{\Nt} \Bigg[ \E\bigg\{\eta_{[n],k} \Big| \mathbf{g}_{[n],k}^\top {\pmb{\varphi}_{[n],k}^\ast} \Big|^2 \int_0^T S_{[n],k}(t)^2 dt \bigg\}\nonumber\\
&+ \E\Bigg\{\Bigg|\sum_{j=1; j\neq k}^K \! \sqrt{\eta_{[n],j}}  \mathbf{g}_{\{[n],j\},k}^\top {\pmb{\varphi}_{[n],j}^\ast} \Bigg|^2 \int_0^T S_{[n],j}(t)^2 dt \Bigg\}\Bigg],\label{eq:harveat_energy_chirp_sensork_ecsi2}
\end{align}
which follows by invoking relations \eqref{eq:xi-chirp_cross-corr} and \eqref{eq:xi-chirp_uncorrelated} above. By using the fact that the chirp waveform has an average power of unity, we obtain 
\begin{equation}
\E\bigg\{\int_0^T \left[S_{[n],i}(t)^2 \right]_{\text{LPF}} dt\bigg\} = \E\bigg\{\sum_{l=1}^\xi d^2_{[n],i,l}\bigg\} T = \xi T,\label{eq:s}
\end{equation}
where $i\in \{k,j\}$ and $\E\{d^2_{[n],i,l}\} = 1$, since $d_{n,i,l} \sim \mathcal{N}(0,1)$. Thus, \eqref{eq:harveat_energy_chirp_sensork_ecsi2} simplifies to
\begin{align}
\mathcal{E}_1 = \varepsilon_2 \Pt &\xi T \sum_{n=1}^{\Nt} \Bigg[\eta_{[n],k} \underbrace{\E\bigg\{\Big| \mathbf{g}_{[n],k}^\top {\pmb{\varphi}_{[n],k}^\ast} \Big|^2 \bigg\}}_{\mathcal{E}_{11}} \nonumber\\
&+ \underbrace{\E\bigg\{\bigg|\sum_{j=1, j\neq k}^K \sqrt{\eta_{[n],j}} \mathbf{g}_{\{[n],j\},k}^\top {\pmb{\varphi}_{[n],j}^\ast} \bigg|^2 \bigg\}}_{\mathcal{E}_{12}} \!\Bigg].\label{eq:harveat_energy_chirp_sensork_ecsi3}
\end{align}
\begin{figure*}[t]\setcounter{equation}{51}
	\vspace{-2mm}
	{\begin{align}
		\E \left\{X_{[n],k} \right\} &= \frac{N!}{(n-1)! (N-n)!} \int_0^\infty \left(\frac{\gamma(M,x_{n,k})}{\Gamma(M)}\right)^{N-n} \left(1-\frac{\gamma(M,x_{n,k})}{\Gamma(M)}\right)^{n-1} \frac{{x^M_{n,k}}}{\Gamma(M)} e^{-x_{n,k}} dx_{n,k}\nonumber\\
		&= \frac{N!}{(n-1)! (N-n)! } \sum_{l=0}^{n-1} (-1)^l \binom{n-1}{l}(\Gamma(M))^{n-N-l-1} \varLambda_1(l).\label{eq:harveat_energy_chirp_y52}
		\end{align}}\vspace{-1mm}\setcounter{equation}{53}
	\begin{align}
	\E \left\{{|\widetilde{h}_{[n],k}|}^2 \right\} = \widetilde{\varUpsilon}^{[n]}_k &= \frac{N!}{(n-1)! (N-n)!} \int_0^\infty x_{n,k}  e^{-n x_{n,k}} (1-e^{-x_{n,k}})^{N-n} dx_{n,k} \nonumber\\
	&= \frac{N!}{(n-1)! (N-n)!} \sum_{l=0}^{N-n} \binom{N-n}{l} \frac{(-1)^{N-n+l}}{(N-l)^2}.\label{eq:term_chirp_ecsi_18}
	\end{align} \setcounter{equation}{62}
	\begin{align}
	\E \big\{z_{[n],k}^4\big\} &= \Pt^2 \eta^2_{[n],k} \E\Bigg\{\Bigg| \frac{\mathbf{g}_{[n],k}^\top \widehat{\mathbf{g}}_{[n],k}^\ast}{\big\|{\widehat{\mathbf{g}}_{[n],k} }\big\|} \Bigg|^4 \Bigg\} = \Pt^2 \eta^2_{[n],k} \E\Bigg\{\Bigg|\frac{\big(\widehat{\mathbf{g}}_{[n],k}^\top\widehat{\mathbf{g}}_{[n],k}^\ast \big) - \big(\widetilde{\mathbf{g}}_{[n],k}^\top \widehat{\mathbf{g}}_{[n],k}^\ast \big) }{\big\|{\widehat{\mathbf{g}}_{[n],k} }\big\|} \Bigg|^4 \Bigg\} \nonumber\\
	&= P^2_\text{tx} \eta^2_{\text{c}_{[n],k}} \Bigg[\mathbb{E}\Bigg\{ \Bigg( \frac{ \widehat{\mathbf{g}}_{[n],k}^{\text{T}}\widehat{\mathbf{g}}_{[n],k}^\ast }{\big\|{\widehat{\mathbf{g}}_{[n],k} }\big\|} \Bigg)^4 \Bigg\} + \mathbb{E}\Bigg\{ \Bigg( \frac{ \widetilde{\mathbf{g}}_{[n],k}^{\text{T}} \widehat{\mathbf{g}}_{[n],k}^\ast }{\big\|{\widehat{\mathbf{g}}_{[n],k} }\big\|} \Bigg)^4 \Bigg\} + 6\mathbb{E}\Bigg\{ \frac{ \big(\widehat{\mathbf{g}}_{[n],k}^{\text{T}}\widehat{\mathbf{g}}_{[n],k}^\ast \big)^2 \big(\widetilde{\mathbf{g}}_{[n],k}^{\text{T}} \widehat{\mathbf{g}}_{[n],k}^\ast \big)^2 }{\big\|{\widehat{\mathbf{g}}_{[n],k} }\big\|^4} \Bigg\} \Bigg] \nonumber\\
	&\approx \Pt^2 \eta^2_{[n],k} \bigg[\gamma^2_k \E\Big\{\big\|{\widehat{\mathbf{h}}_{[n],k}}\big\|^4 \Big\} +  6\E\Big\{\widehat{\mathbf{g}}_{[n],k}^\top \big(\widetilde{\mathbf{g}}_{[n],k}^\ast \widetilde{\mathbf{g}}_{[n],k}^\top \big) \widehat{\mathbf{g}}_{[n],k}^\ast \Big\} %\nonumber\\
	+ \bigg(\! \E\bigg\{\frac{ \widehat{\mathbf{g}}_{[n],k}^\top \big(\widetilde{\mathbf{g}}_{[n],k}^\ast \widetilde{\mathbf{g}}_{[n],k}^\top \big) \widehat{\mathbf{g}}_{[n],k}^\ast }{\big\|{\widehat{\mathbf{g}}_{[n],k}}\big\|^2} \bigg\} \bigg)^{2}\bigg]. \label{eq:harveat_energy_chirp_y49}
	\end{align}		
	\noindent\makebox[\linewidth]{\rule{18.3cm}{.4pt}}\vspace{-5mm}
\end{figure*}\setcounter{equation}{48}
From \eqref{eq:estimated_channel_vector}, we have $\mathbf{g}_{[n],k} = \widehat{\mathbf{g}}_{[n],k} - \widetilde{\mathbf{g}}_{[n],k}$, 
where $\widehat{\mathbf{g}}_{[n],k} = \sqrt{\gamma_k} \widehat{\mathbf{h}}_{[n],k}$, $\widetilde{\mathbf{g}}_{[n],k} = \sqrt{\beta_k-\gamma_k} \widetilde{\mathbf{h}}_{[n],k}$, and the channel estimation error $\widetilde{\mathbf{g}}_{[n],k}$ is uncorrelated to the estimated channel $\widehat{\mathbf{g}}_{[n],k}$. Now, $\mathcal{E}_{11}$ in \eqref{eq:harveat_energy_chirp_sensork_ecsi3} is expressed as
\begin{equation}
\mathcal{E}_{11} = \gamma_k \E\left\{\big\|{\widehat{\mathbf{h}}_{[n],k} }\big\|^2 \right\} + \E\Bigg\{ \frac{\widehat{\mathbf{g}}_{[n],k}^\top \big(\widetilde{\mathbf{g}}_{[n],k}^\ast \widetilde{\mathbf{g}}_{[n],k}^\top \big) \widehat{\mathbf{g}}_{[n],k}^\ast }{\big\|{\widehat{\mathbf{g}}_{[n],k} }\big\|^2} \Bigg\}.\label{eq:harveat_energy_chirp_y22}
\end{equation}
Since $\widehat{h}_{n,k}^m$ is a CSCG random variable, $X_{n,k} = \big\|{\widehat{\mathbf{h}}_{n,k}}\big\|^2 = \sum_{m=1}^M {|\widehat{h}_{n,k}^m|}^2$ is a gamma random variable with shape parameter $M$ and rate parameter one. Therefore,
\begin{equation}
\E\{X_{[n],k}\} = \int_0^\infty x_{n,k} f_{X_{[n],k}}(x_{n,k}) dx_{n,k},\label{eq:term_chirp_ecsi_15}
\end{equation}
where, based on order statistics~\cite{Nadarajah}, the PDF of the $n^{\text{th}}$ largest random variable is given by 
\begin{align}
f_{X_{[n], k}}(x) = &\frac{N!}{(n-1)! (N-n)!} (F_{X_{n,k}}(x))^{N-n} \nonumber\\
&\hspace{15mm}\times (1-F_{X_{n,k}}(x))^{n-1} f_{X_{n,k}}(x),\label{eq:order_stat_6}
\end{align}
with $f_{X_{n,k}}(x)$ and $F_{X_{n,k}}(x)$ being the PDF and CDF of $X_{n,k}$, respectively \cite{Papoulis}. By substituting \eqref{eq:order_stat_6} in \eqref{eq:term_chirp_ecsi_15} and by using the distributions of a gamma random variable, we obtain $\E \left\{X_{[n],k} \right\}$ in \eqref{eq:harveat_energy_chirp_y52} where $\varLambda_1(l)$ can be expressed using the Lauricella function of type A \cite[Eq. (7)]{Nadarajah} as in \eqref{eq:thrm1_Omegaklambda}. From (\ref{eq:harveat_energy_chirp_y22}), $\E \big\{\widetilde{\mathbf{g}}_{[n],k}^\ast \widetilde{\mathbf{g}}_{[n],k}^\top \big\}$ is expressed as
\setcounter{equation}{52}
\begin{align}
&\E \left\{\widetilde{\mathbf{g}}_{[n],k}^\ast \widetilde{\mathbf{g}}_{[n],k}^\top \right\} \nonumber\\
&= \left(\beta_k-\gamma_k \right) \E \left\{ \text{diag} \left\{ {|\widetilde{h}_{[n],k}^1|}^2 , {|\widetilde{h}_{[n],k}^2|}^2 , \ldots, {|\widetilde{h}_{[n],k}^M|}^2 \right\} \right\},\label{eq:term_chirp_ecsi_17}
\end{align}
where ${|\widetilde{h}^m_{n,k}|}^2$ is an exponential random variable with parameter one. Similar to above, we evaluate $\E \left\{\!{|\widetilde{h}_{[n],k}|}^2 \right\} \!= \widetilde{\varUpsilon}^{[n]}_k$ in \eqref{eq:term_chirp_ecsi_18} 
where the integral is evaluated using \cite[(3.432.1)]{Gradshteyn}. Hence, substituting \eqref{eq:harveat_energy_chirp_y52} and \eqref{eq:term_chirp_ecsi_18} in \eqref{eq:harveat_energy_chirp_y22}, we can write
\setcounter{equation}{54}
\begin{align}
\mathcal{E}_{11} &= \gamma_k \E\{X_{[n],k}\} + (\beta_k-\gamma_k) \E \big\{{|\widetilde{h}_{[n],k}|}^2 \big\}  \nonumber\\
&= \varOmega^{[n]}_{k,1} + \varUpsilon^{[n]}_k,
\end{align}
where $\varUpsilon^{[n]}_k$ and $\varOmega^{[n]}_{k,1}$ are given by \eqref{eq:upsilon} and \eqref{eq:omega}, respectively.
Next, by expanding $\mathcal{E}_{12}$ of \eqref{eq:harveat_energy_chirp_sensork_ecsi3}, we have
\begin{align}
&\mathcal{E}_{12} = \sum_{j=1; j\neq k}^K \eta_{[n],j} \E \bigg\{\frac{\widehat{\mathbf{g}}_{[n],j}^\top (\mathbf{g}_{\{[n],j\},k}^\ast \mathbf{g}_{\{[n],j\},k}^\top) \widehat{\mathbf{g}}_{[n],j}^\ast}{\|{\widehat{\mathbf{g}}_{[n],j}}\|^2} \bigg\} \nonumber\\
&+ \sum_{\substack{j=1\\ j\neq k}}^K \sum_{\substack{i=1\\ i\neq j,k}}^K \sqrt{\eta_{[n],j} \eta_{[n],i}} \E \bigg\{\frac{\widehat{\mathbf{g}}_{[n],j}^\top (\mathbf{g}_{\{[n],j\},k}^\ast \mathbf{g}_{\{[n],i\},k}^\top) \widehat{\mathbf{g}}_{[n],i}^\ast }{\|{\widehat{\mathbf{g}}_{[n],j}}\| \|{\widehat{\mathbf{g}}_{[n],i}}\|} \bigg\}.\label{eq:term_chirp_ecsi_20}
\end{align}
All channel realizations are independent to each other and the ordered subbands for the $j^\text{th}$ user are not the same as that of the ordered subbands for the $k^\text{th}$ user. Thus, $\E\big\{\mathbf{g}_{\{[n],j\},k}^\ast \mathbf{g}_{\{[n],j\},k}^\top\big\} = \beta_k \mathbf{I}_M$. Due to the independent channels, we also have $\E \big\{ \mathbf{g}_{\{[n],j\},k}^\ast \mathbf{g}_{\{[n],i\},k}^\top \big\}=0$. Moreover, since the index of the $n^\text{th}$ best subband of the $j^\text{th}$ user may be different from the index of the $n^\text{th}$ best subband of the $i^\text{th}$ user and since the channels corresponding to different users are independent, $\E \big\{\widehat{\mathbf{g}}_{[n],j}^\top \widehat{\mathbf{g}}_{[n],i}^\ast \big\}=0$. Therefore, the second term of \eqref{eq:term_chirp_ecsi_20} is zero. By using these observations, \eqref{eq:term_chirp_ecsi_20} simplifies to
\begin{equation}
\mathcal{E}_{12} = \beta_k \sum_{j=1; j\neq k}^K \eta_{[n],j}.
\label{eq:term_chirp_ecsi_21}
\end{equation}
Substituting the final expressions for $\mathcal{E}_{11}$ and $\mathcal{E}_{12}$ in \eqref{eq:harveat_energy_chirp_sensork_ecsi3}, $\mathcal{E}_1$ can be expressed as the first term of \eqref{eq:thrm1_harveat_energy_chirp_sensork}.

Next, for $m=3$, $y_k(t)^3$ is written as
\begin{equation}
y_k(t)^3 = v_k(t)^3 + v_j(t)^3 + 3v_k(t)^2 v_j(t) + 3v_k(t) v_j(t)^2.\label{eq:harveat_energy_chirp_y3}
\end{equation}
Since $S_{n,k}(t)$ is a periodic and even function, we get $\varepsilon_3\E \{\int_0^T [y_k(t)^3]_{\text{LPF}} dt\} = 0$.

Finally, for $m=4$, $y_k(t)^4$ is given by
\begin{align}
y_k(t)^4 &= v_k(t)^4 + v_j(t)^4 + 4(v_k(t)^3 v_j(t) + v_k(t) v_j(t)^3) \nonumber\\
&\hspace{40mm} + 6 v_k(t)^2 v_j(t)^2 \nonumber\\
&= v_k(t)^4 + v_j(t)^4 + 6 v_k(t)^2 v_j(t)^2, \label{eq:harveat_energy_chirp_y41}
\end{align}
where the expectation of the time-average for the third term is zero as $S_{n,k}(t)$ is a periodic and even function. Accordingly, $\mathcal{E}_2 = \varepsilon_4\E \{\int_0^T [y_k(t)^4]_{\text{LPF}} dt\}$ is evaluated as
\begin{equation}
\mathcal{E}_2 = \varepsilon_4 \E \bigg\{\int_0^T \big[v_k(t)^4 + v_j(t)^4 + 6 v_k(t)^2 v_j(t)^2\big]_{\text{LPF}} dt \bigg\}.\label{eq:harveat_energy_chirp_sensork_y47}
\end{equation}

To evaluate $\E \big\{\int_0^T [v_k(t)^4]_{\text{LPF}} dt \big\}$, observe that the expectation of the time-average for some terms is zero and some terms are filtered out by the LPF. Therefore, by considering the effective non-zero terms, $\E\{[v_k(t)^4]_{\text{LPF}}\}$ is given by
\begin{align}
&\E \bigg\{\!\int_0^T \!\Big[v_k(t)^4\Big]_{\text{LPF}} dt\bigg\} \!=\! \sum_{n=1}^{\Nt} \!\E\bigg\{\!\int_0^T \!\Big[z_{[n],k}^4 S_{[n],k}(t)^4 \Big]_{\text{LPF}} dt\bigg\}\!\nonumber\\
&+ 3\sum_{n \neq u}^{\Nt} \E\bigg\{\!\int_0^T \!\Big[z_{[n],k}^2 z_{[u],k}^2 S_{[n],k}(t)^2 S_{[u],k}(t)^2\Big]_{\text{LPF}} dt\bigg\}\!,\!\!\label{eq:harveat_energy_chirp_y45}
\end{align}
where $z_{[n],k} = \sqrt{\Pt\eta_{[n],k}} \mathbf{g}_{[n],k}^\top {\pmb{\varphi}_{[n],k}^\ast}$. We have that
\begin{align}
\E\bigg\{& \int_0^T \big[S_{[n],k}(t)^4 \big]_{\text{LPF}} dt\bigg\} \nonumber\\
&= \E\bigg\{\frac{3}{2}\sum_{l=1}^\xi d^4_{[n],k,l} + 6 \sum_{i<j}^\xi d^2_{[n],k,i} d^2_{[n],k,j}\bigg\} T \nonumber\\
&= \left[\frac{9}{2}\xi + 6 \binom{\xi}{2}\right] T,
\end{align}
which follows from $\E\{d^4_{[n],k,l}\} = 3$ and $\E\{d^2_{[n],k,l}\} = 1$. Also, $\E \big\{z_{[n],k}^4\big\}$ is evaluated in \eqref{eq:harveat_energy_chirp_y49}, where the last line follows from the fact that $\E\{X^2\} \geq (\E\{X\})^2$. Recall that $X_{[n],k} = \big\|{\widehat{\mathbf{h}}_{[n],k} }\big\|^2$. Then, by using similar steps as for \eqref{eq:harveat_energy_chirp_y52}, we have $\E\big\{X_{[n],k}^2 \big\} = \E\big\{ \big\|{\widehat{\mathbf{h}}_{[n],k}}\big\|^4 \big\}$ and is given in \eqref{eq:harveat_energy_chirp_y531}, where $\varLambda_2(l)$ is given in \eqref{eq:thrm1_Omegaklambda}. Following the fact that the channel estimation error $\widetilde{\mathbf{g}}_{[n],k}$ is uncorrelated to the estimated channel $\widehat{\mathbf{g}}_{[n],k}$, the second inner term of \eqref{eq:harveat_energy_chirp_y49} is evaluated with the help of \eqref{eq:harveat_energy_chirp_y22}, \eqref{eq:harveat_energy_chirp_y52} and \eqref{eq:term_chirp_ecsi_18} as follows
\setcounter{equation}{64}
\begin{equation}	
\E\Big\{\widehat{\mathbf{g}}_{[n],k}^\top \big(\widetilde{\mathbf{g}}_{[n],k}^\ast \widetilde{\mathbf{g}}_{[n],k}^\top \big) \widehat{\mathbf{g}}_{[n],k}^\ast \Big\} = \varOmega^{[n]}_{k,1} \varUpsilon^{[n]}_k,\label{eq:harveat_energy_chirp_y51}
\end{equation}
where $\varUpsilon^{[n]}_k$ and $\varOmega^{[n]}_{k,1}$ are given by \eqref{eq:upsilon} and \eqref{eq:omega}, respectively. The last inner term of \eqref{eq:harveat_energy_chirp_y49} is evaluated using \eqref{eq:term_chirp_ecsi_17} and \eqref{eq:term_chirp_ecsi_18} as
\begin{equation}
\Bigg(\E\Bigg\{\frac{\widehat{\mathbf{g}}_{[n],k}^\top \big(\widetilde{\mathbf{g}}_{[n],k}^\ast \widetilde{\mathbf{g}}_{[n],k}^\top\big) \widehat{\mathbf{g}}_{[n],k}^\ast}{\big\|{\widehat{\mathbf{g}}_{[n],k}}\big\|^2}\Bigg\}\Bigg)^2 = \big(\varUpsilon^{[n]}_k\big)^2.\label{eq:harveat_energy_chirp_y53}
\end{equation}
Next, using \eqref{eq:harveat_energy_chirp_y22}, $\E\big\{z_{[n],k}^2 z_{[u],k}^2\big\}$ is evaluated as 
\begin{align}
\E\big\{z_{[n],k}^2 z_{[u],k}^2\big\} &= \Pt^2 \eta_{[n],k} \big(\varOmega^{[n]}_{k,1} + \varUpsilon^{[n]}_k \big)\eta_{[u],k} \big(\varOmega^{[u]}_{k,1} + \varUpsilon^{[u]}_k\big),\label{eq:harveat_energy_chirp_y54}
\end{align}
and from \eqref{eq:s} we have 
\begin{equation}
	\E\bigg\{\int_0^T \Big[S_{[n],k}(t)^2 S_{[u],k}(t)^2 \Big]_{\text{LPF}} dt\bigg\} = \xi^2 T.\label{eq:s2}
\end{equation} 
\begin{figure*}[t]\setcounter{equation}{63}
		\begin{align}
	\E\big\{ \big\|{\widehat{\mathbf{h}}_{[n],k}}\big\|^4 \big\} &= \frac{N! ~ \sum\limits_{l= 0}^{n-1} (-1)^l {\! n-1 \!\choose l\!} (\Gamma(M))^{n-N-l-1}}{(n-1)! (N-n)! } 
	\int_{0}^{\infty}\! {x^{M+1}_{n,k}} e^{-x_{n,k}} \left(\gamma(M,x_{n,k}) \right)^{l+N-n} dx_{n,k} \nonumber\\
	&= \frac{N!}{(n-1)! (N-n)!} \sum_{l=0}^{n-1} (-1)^l \binom{n-1}{l} (\Gamma(M))^{n-N-l-1} \varLambda_2(l).\label{eq:harveat_energy_chirp_y531}
	\end{align} \vspace{-0.5mm}\setcounter{equation}{70}
	{\begin{align}
		\E\Bigg\{\int_0^T \!\Bigg[ \sum_{\substack{j=1\\ j\neq k}}^K z_{\{[n],k\},j}(t)\Bigg]^4_{\text{LPF}} dt\Bigg\} &= \sum_{\substack{j=1\\ j\neq k}}^K \E\bigg\{\int_0^T z_{\{[n],k\},j}(t)^4_{\text{LPF}} dt\bigg\} + 3\sum_{\substack{j\neq p\\ j\neq k; p\neq k}}^K \E \bigg\{\int_0^T \!\Big[z_{\{[n],k\},j}(t)^2 z_{\{[n],k\},p}(t)^2\Big]_{\text{LPF}} dt \bigg\} \nonumber\\
		&\approx \Pt^2 \beta_k^2 \left[\frac{9}{2}\xi + 6 \binom{\xi}{2}\right] T \sum_{j=1; j\neq k}^K \eta_{[n],j}^2 + 3 \Pt^2 \beta_k^2 \xi^2 T \sum_{\substack{j\neq p; j\neq k; p\neq k}}^K \eta_{[n],j}\eta_{[n],p}.\label{eq:harveat_energy_chirp_y56}
		\end{align}}%\vspace{-1mm}%\setcounter{equation}{34}
	\noindent\makebox[\linewidth]{\rule{18.3cm}{.4pt}}\vspace{-5mm}
\end{figure*}\setcounter{equation}{68}
Similarly, for $\E\big\{\int_0^T [v_j(t)^4]_{\text{LPF}} dt\big\}$ we can write
\begin{align}
&\!\E\bigg\{\!\!\int_0^T \big[v_j(t)^4\big]_{\text{LPF}} dt\!\bigg\} \!=\! \sum_{n=1}^{\Nt} \!\E\Bigg\{\!\!\int_0^T\!\!\Bigg(\!\sum_{j=1; j\neq k}^K \!z_{\{[n],k\},j}(t)\!\Bigg)^{\!4} dt \!\Bigg\}\!\nonumber\\
&+ 3\sum_{n\neq u}^{\Nt} \!\E\Bigg\{\!\int_0^T \!\Bigg(\sum_{\substack{j=1\\ j\neq k}}^K z_{\{[n],k\},j}(t)\!\Bigg)^2 \Bigg(\sum_{\substack{j=1\\ j\neq k}}^K z_{\{[u],k\},j}(t)\!\Bigg)^2 dt\Bigg\},\label{eq:harveat_energy_chirp_y46}
\end{align}
where
\begin{align}
z_{\{[n],k\},j}(t) = \sqrt{\Pt \eta_{[n],j}} \mathbf{g}_{\{[n],j\},k}^\top {\pmb{\varphi}_{[n],j}^\ast} S_{[n],j}(t).
\end{align}
Moreover, considering only the effective non-zero terms, we can write $\E\Big\{\int_0^T \Big[ \sum_{\substack{j=1; j\neq k}}^K z_{\{[n],k\},j}(t)\Big]^4_{\text{LPF}} dt\Big\}$ as in \eqref{eq:harveat_energy_chirp_y56} which can be evaluated by following similar steps as the derivation of \eqref{eq:term_chirp_ecsi_20} and \eqref{eq:harveat_energy_chirp_y49}. For the second term of \eqref{eq:harveat_energy_chirp_y46}, we can use the result in \eqref{eq:term_chirp_ecsi_21} and write
\setcounter{equation}{71}
\begin{align}
&\E\Bigg\{\int_0^T\!\Bigg(\sum_{\substack{j=1; j\neq k}}^K z_{\{[n],k\},j}(t)\Bigg)^2 \Bigg(\sum_{\substack{j=1; j\neq k}}^K z_{\{[u],k\},j}(t)\Bigg)^2dt\Bigg\}\nonumber\\
&= T \Bigg(\xi \Pt \beta_k\sum_{\substack{j=1; j\neq k}}^K \eta_{[n],j}\Bigg)\Bigg(\xi \Pt \beta_k \sum_{\substack{j=1; j\neq k}}^K \eta_{[u],j}\Bigg). 
\label{eq:harveat_energy_chirp_y55}
\end{align} 

For the third term of \eqref{eq:harveat_energy_chirp_sensork_y47}, with the help of the derivation of \eqref{eq:harveat_energy_chirp_sensork_ecsi2}, we can evaluate  $\E\big\{\int_0^T [v_k(t)^2 v_j(t)^2]_{\text{LPF}} dt\big\}$ as 
\begin{align}
%&\E\{\int_0^T v_k(t)^2 v_j(t)^2 dt \} = 
&\E\bigg\{\int_0^T \bigg[\Pt^2 \sum_{n=1}^{\Nt} \eta_{[n],k} \Big|\mathbf{g}_{[n],k}^\top {\pmb{\varphi}_{[n],k}^\ast} \Big|^2 S_{[n],k}(t)^2 \nonumber\\
&\times \bigg(\sum_{n=1}^{\Nt}\sum_{j=1; j\neq k}^K \! \eta_{[n],j} \Big|\mathbf{g}_{\{[n],j\},k}^\top {\pmb{\varphi}_{[n],j}^\ast}\Big|^2 S_{[n],j}(t)^2\bigg) \bigg]_{\text{LPF}} dt\bigg\}\nonumber\\
&\approx\! \Pt^2 \beta_k \xi^2 T\sum_{n=1}^{\Nt} \eta_{[n],k} \Big(\! \varOmega^{[n]}_{k,1} + \varUpsilon^{[n]}_k\Big) \sum_{u=1}^{\Nt} \sum_{j=1; j\neq k}^K \! \eta_{[u],j},\!\label{eq:harveat_energy_chirp_y60} %(1-\eta_{[u],k})
\end{align}
where the terms with higher frequencies are neglected following LPF.

Finally, by substituting the derive expressions for $\E\{v_k(t)^4\}$, $\E\{v_j(t)^4\}$ and $\E\{v_k(t)^2 v_j(t)^2\}$ in \eqref{eq:harveat_energy_chirp_sensork_y47} and after several algebraic manipulations, we end up with the second term of \eqref{eq:thrm1_harveat_energy_chirp_sensork}, and the theorem is proven.

\bibliographystyle{IEEEtran}
\bibliography{Bibtex/bibJournalList,Bibtex/book,Bibtex/references}

\end{document}